\documentclass[12pt]{article}

\usepackage [latin1]{inputenc}

\usepackage{lscape}
\usepackage{graphics}
\usepackage{epsfig}
\usepackage{graphicx}
\usepackage{color}
\usepackage{amsmath}

\begin{document}
\baselineskip=24pt

\newcommand{\D}{\displaystyle} 
\newcommand{\T}{\textstyle} 
\newcommand{\SC}{\scriptstyle} 
\newcommand{\SSC}{\scriptscriptstyle} 

\newcommand{\be}{\begin{eqnarray}}
\newcommand{\ee}{\end{eqnarray}}

\definecolor{yellow}{rgb}{0.95,0.75,0.1}
\definecolor{orange}{rgb}{0.95,0.4,0.1}
\definecolor{red}{rgb}{1,0,0}
\definecolor{green}{rgb}{0,1,0}
\definecolor{blue}{rgb}{0,0.5,1}

\definecolor{lblue}{rgb}{0,0.8,1}
\definecolor{dblue}{rgb}{0,0,1}
\definecolor{dgreen}{rgb}{0,0.65,0}
\definecolor{lila}{rgb}{0.8,0,0.8}
\definecolor{violet}{rgb}{1,0,0.9}
\definecolor{grey}{rgb}{0.3,0.3,0.3}

\definecolor{contoura}{rgb}{0,0,1}
\definecolor{contourb}{rgb}{0,1,1}
\definecolor{contourc}{rgb}{0,1,0}
\definecolor{contourd}{rgb}{0.95,0.75,0.1}
\definecolor{contoure}{rgb}{1,0,0}
\definecolor{contourf}{rgb}{1,0,1}

\newcommand\cred[1]{\textcolor{red}{#1}}
\newcommand\cblue[1]{\textcolor{blue}{#1}}
\newcommand\ccyan[1]{\textcolor{contourb}{#1}}
\newcommand\cgreen[1]{\textcolor{green}{#1}}
\newcommand\cyellow[1]{\textcolor{yellow}{#1}}
\newcommand\cviolet[1]{\textcolor{violet}{#1}}
\newcommand\cmagenta[1]{\textcolor{magenta}{#1}}
\newcommand\cgrey[1]{\textcolor{grey}{#1}}

\newcommand\cconta[1]{\textcolor{contoura}{#1}}
\newcommand\ccontb[1]{\textcolor{contourb}{#1}}
\newcommand\ccontc[1]{\textcolor{contourc}{#1}}
\newcommand\ccontd[1]{\textcolor{contourd}{#1}}
\newcommand\cconte[1]{\textcolor{contoure}{#1}}
\newcommand\ccontf[1]{\textcolor{contourf}{#1}}

\newcommand {\black} {\color{black}}
\newcommand {\violet}{\color{violet}}
\newcommand {\blue} {\color{blue}}
\newcommand {\dblue} {\color{dblue}}
\newcommand {\cyan} {\color{cyan}}
\newcommand {\lila} {\color{lila}}
\newcommand {\yellow} {\color{yellow}}
\newcommand {\green} {\color{green}}
\newcommand {\dgreen} {\color{dgreen}}
\newcommand {\magenta} {\color{magenta}}
\newcommand {\red} {\color{red}}
\newcommand {\orange} {\color{orange}}

\def\AJ{{\it Astron. J.} }
\def\ARAA{{\it Annual Rev. of Astron. \& Astrophys.} }
\def\ARNPS{{\it Annual Rev. of Nucl. \& Part. Sci. } }
\def\ApJ{{\it Astrophys. J.} }
\def\ApJL{{\it Astrophys. J. Letters} }
\def\ApJS{{\it Astrophys. J. Suppl.} }
\def\ApP{{\it Astropart. Phys.} }
\def\AA{{\it Astron. \& Astroph.} }
\def\AAR{{\it Astron. \& Astroph. Rev.} }
\def\AAL{{\it Astron. \& Astroph. Letters} }
\def\AASu{{\it Astron. \& Astroph. Suppl.} }
\def\AN{{\it Astron. Nachr.} }
\def\ASR{{\it Adv. in Space Res.} }
\def\EPJC{{\it Eur. Phys. Journ.} {\bf C} }
\def\IJMP{{\it Int. J. of Mod. Phys.} }
\def\JCAP{{\it J. of Cosmol. and Astrop. Phys.} }
\def\JGR{{\it Journ. of Geophys. Res.}}
\def\JHEP{{\it Journ. of High En. Phys.} }
\def\JPhG{{\it Journ. of Physics} {\bf G} }
\def\CQG{{\it Class. Quant. Grav. } }
\def\MNRAS{{\it Month. Not. Roy. Astr. Soc.} }
\def\Nature{{\it Nature} }
\def\NewAR{{\it New Astron. Rev.} }
\def\NewA{{ New Astron.} }
\def\NIMPA{{\it Nucl. Instr. Meth. Phys. Res.}{\bf A} }
\def\PASP{{\it Publ. Astron. Soc. Pac.}}
\def\PhFl{{\it Phys. of Fluids} }
\def\PLB{{\it Phys. Lett.}{\bf B} }
\def\PR{{\it Phys. Rev.} }
\def\PRD{{\it Phys. Rev.} {\bf D} }
\def\PRL{{\it Phys. Rev. Letters} }
\def\PRX{{\it Phys. Rev. }{\bf X} }
\def\RMP{{\it Rev. Mod. Phys.} }
\def\RPP{{\it Rep. Pro.Phys.} }
\def\Science{{\it Science}}
\def\ZfA{{\it Zeitschr. f{\"u}r Astrophys.} }
\def\ZfN{{\it Zeitschr. f{\"u}r Naturforsch.} }
\def\etal{{\it et al.}}

\hyphenation{mono-chro-matic  sour-ces  Wein-berg
chang-es Strah-lung dis-tri-bu-tion com-po-si-tion elec-tro-mag-ne-tic
ex-tra-galactic ap-prox-i-ma-tion nu-cle-o-syn-the-sis re-spec-tive-ly
su-per-nova su-per-novae su-per-nova-shocks con-vec-tive down-wards
es-ti-ma-ted frag-ments grav-i-ta-tion-al-ly el-e-ments me-di-um
ob-ser-va-tions tur-bul-ence sec-ond-ary in-ter-action
in-ter-stellar spall-ation ar-gu-ment de-pen-dence sig-nif-i-cant-ly
in-flu-enc-ed par-ti-cle sim-plic-i-ty nu-cle-ar smash-es iso-topes
in-ject-ed in-di-vid-u-al nor-mal-iza-tion lon-ger con-stant
sta-tion-ary sta-tion-ar-i-ty spec-trum pro-por-tion-al cos-mic
re-turn ob-ser-va-tion-al es-ti-mate switch-over grav-i-ta-tion-al
super-galactic com-po-nent com-po-nents prob-a-bly cos-mo-log-ical-ly
Kron-berg Berk-huij-sen}
\def\simle{\lower 2pt \hbox {$\buildrel < \over {\scriptstyle \sim }$}}
\def\simge{\lower 2pt \hbox {$\buildrel > \over {\scriptstyle \sim }$}}
\def\intunits{{\rm s}^{-1}\,{\rm sr}^{-1} {\rm cm}^{-2}}

\def\sun{\hbox{$\odot$}}
\def\la{\mathrel{\mathchoice {\vcenter{\offinterlineskip\halign{\hfil
$\displaystyle##$\hfil\cr<\cr\sim\cr}}}
{\vcenter{\offinterlineskip\halign{\hfil$\textstyle##$\hfil\cr
<\cr\sim\cr}}}
{\vcenter{\offinterlineskip\halign{\hfil$\scriptstyle##$\hfil\cr
<\cr\sim\cr}}}
{\vcenter{\offinterlineskip\halign{\hfil$\scriptscriptstyle##$\hfil\cr
<\cr\sim\cr}}}}}
\def\ga{\mathrel{\mathchoice {\vcenter{\offinterlineskip\halign{\hfil
$\displaystyle##$\hfil\cr>\cr\sim\cr}}}
{\vcenter{\offinterlineskip\halign{\hfil$\textstyle##$\hfil\cr
>\cr\sim\cr}}}
{\vcenter{\offinterlineskip\halign{\hfil$\scriptstyle##$\hfil\cr
>\cr\sim\cr}}}
{\vcenter{\offinterlineskip\halign{\hfil$\scriptscriptstyle##$\hfil\cr
>\cr\sim\cr}}}}}
\def\degr{\hbox{$^\circ$}}
\def\arcmin{\hbox{$^\prime$}}
\def\arcsec{\hbox{$^{\prime\prime}$}}
\def\utw{\smash{\rlap{\lower5pt\hbox{$\sim$}}}}
\def\udtw{\smash{\rlap{\lower6pt\hbox{$\approx$}}}}
\def\fd{\hbox{$.\!\!^{\rm d}$}}
\def\fh{\hbox{$.\!\!^{\rm h}$}}
\def\fm{\hbox{$.\!\!^{\rm m}$}}
\def\fs{\hbox{$.\!\!^{\rm s}$}}
\def\fdg{\hbox{$.\!\!^\circ$}}
\def\farcm{\hbox{$.\mkern-4mu^\prime$}}
\def\farcs{\hbox{$.\!\!^{\prime\prime}$}}
\def\fp{\hbox{$.\!\!^{\scriptscriptstyle\rm p}$}}
\def\baselinestretch{1.5}
\def\gsim{\stackrel{>}{\sim}}
\def\lsim{\stackrel{<}{\sim}}
\def\beq{\begin{equation}}
\def\eeq{\end{equation}}
\def\ol{\overline}

\centerline{\bf Frontiers in Astronomy and Space Sciences}
%
%

\vskip0.5cm

\begin{centering}
{\bf {\large Cosmic ray contributions from rapidly rotating stellar mass black holes: Cosmic Ray GeV to EeV proton and anti-proton sources}}%
\footnote[1]{{\bf Corresponding authors: Athina Meli  and  Peter L. Biermann.} $\dagger$: no longer with us}


{\small
{  M. L. Allen} (Department of Physics \& Astronomy, Washington State University, Pullman, WA 99164, USA); 
{  P. L. Biermann} (i) MPI for Radioastr., Auf dem H{\"u}gel 69, D-53121 Bonn, Germany; ii) Dept. of Phys. \& Astron., U. Alabama, Box 870324, Tuscaloosa, AL 35487-0324, USA); 
{  A. Chieffi} (Istituto Nazionale Di Astrofisica (INAF) $-$ Istituto di Astrofisica e Planetologia Spaziali, Via Fosso del Cavaliere 100, I-00133 Roma, Italy);  
{  R. Chini} (i) Astronomisches Institut, Ruhr$-$Universit\"at Bochum, Universit\"atsstra{\ss}e 150, D-44801 Bochum, Germany; ii) Centrum Astronomiczne im. Mikolaja Kopernika, PAN, Bartycka 18, PL-00-716 Warsaw, Poland; iii) Instituto de Astronom\'{i}a, Universidad Cat\'{o}lica del Norte, Avenida Angamos 0610, Antofagasta, Chile); 
{  D. Frekers} (Institut f{\"u}r Kernphysik, Westf{\"a}lische Wilhelms-Universit{\"a}t M{\"u}nster, D-48149 M{\"u}nster, Germany); 
{  L.{\'A}. Gergely} (i) Univ. Szeged, Department of Theoretical Physics, Szeged, Hungary; ii) HUN-REN (Hungarian Research Network) Wigner Research Centre for Physics, Department of Theoretical Physics, Budapest, Hungary); 
{  Gopal-Krishna} (UM-DAE Centre for Excellence in Basic Sciences,  Vidyanagari, Mumbai-400098, India); 
{  B. Harms$^{\dagger}$} (Dept. of Phys. \& Astron., U. Alabama, Box 870324, Tuscaloosa, AL 35487-0324, USA);  
{  I. Jaroschewski} (Faculty of Physics and Astronomy, Ruhr$-$Universit\"at Bochum, Universit\"atsstra{\ss}e 150, D-44801 Bochum, Germany); %
{  P. S. Joshi} (Int. Center for Space \& Cosmology, Ahmedabad Univ., Ahmedabad 380009, India); 
{  P. P. Kronberg$^{\dagger}$} (Dept. of Phys., Univ. Toronto, 60 St George Street, Toronto, ON M5S 1A7, Canada.); 
{  E. Kun} (Faculty of Physics and Astronomy, Ruhr$-$Universit\"at Bochum, Universit\"atsstra{\ss}e 150, D-44801 Bochum, Germany); 
{  A. Meli} (i) Dep. of Physics, North Carolina A\&T State Univ., Greensboro, NC 27411, USA; ii) Space Sci. \& Techn. for Astrophys. Res. (STAR) Institute, Universit{\'e} de Li{\`e}ge, B-4000 Li{\`e}ge, Belgium);  
{  E.-S. Seo}, (Inst. for Physical Science and Technology, 4254 Stadium Dr, University of Maryland, College Park, MD  20742, USA). 
{  T. Stanev}, (Bartol Res. Inst. and Depart. of Phys. and Astron., Univ. of Delaware, Newark, DE 19716, USA) 
}

manuscript Feb 1, 2024; last revised Sep 25, 2024
\end{centering}

\section{Abstract}
{\small In Radio Super Novae (RSNe) a magnetic field of $(B \, \times \, r) \, = \, 10^{16.0 \pm 0.12} \, {\rm Gauss \, \times \, cm}$  is observed; these are the same numbers for Blue Super Giant (BSG) star explosions as for Red Super Giant (RSG) star explosions, despite their very different wind properties. The EHT data for M87 as well for low power radio galaxies all show consistency with just this value of the quantity $(B \, \times \, r )$, key for angular momentum and energy transport, and can be derived from the radio jet data.  We interpret this as a property of the near surroundings of a black hole (BH) at near maximal rotation, independent of BH mass. In the commonly used green onion model, in which a $2 \, \pi$ flow changes over to a jet flow we interpret this as a wind emanating from the BH/accretion disk system and its surroundings. Near the BH collisions in the wind can produce a large fraction of anti-protons. In this scenario the cosmic Ray (CR) population from the wind/jet is proposed to be visible as EeV protons and anti-protons in the CR data to EeV energy, with a $E^{-7/3}$ spectrum. This can be connected to a concept of inner and outer Penrose zones in the ergo-region. The observed numbers for the magnetic field imply the Planck time as the governing time scale: A BH rotating near maximum can accept a proton per log bin of energy in an extended spectrum with the associated pions every Planck time.}

\section{Introduction: Energetic particles and black holes}

Energetic particles, commonly called Cosmic Ray particles, or short just {\it Cosmic Rays} have been researched since their discovery in 1912 (with a recent review with many references in \cite{ASR18}); further important viewpoints and history are given by Colgate and Yodh \cite{Colgate94,Yodh92,Yodh03,Yodh05,Yodh06}. Cosmic Ray (short CRs) particles have been observed from below GeV, with stellar sources responsible up to a few EeV, as discussed here, and extragalactic sources up to a few hundred EeV. Many of them, both Galactic and extragalactic, can be traced to the activity of black holes.

The various possible sources of CRs were discussed in \cite{ASR18,Gal19}, and earlier papers \cite{CR-I,CR-II,CR-III,CR-IV,Rachen93} with reviews in \cite{Biermann94,Biermann97}. A main distinction, which we have made \cite{CR-IV}, is to differentiate between SNe, that explode into their own wind, wind-SNe, and those that explode into the Interstellar Medium (ISM), ISM-SNe. It is also necessary to sub-divide those two groups: There are Red Super Giant (RSG) stars with slow dense winds, and Blue Super Giant (BSG) stars that explode into tenuous fast winds, heavily enriched in the chemical elements of higher nucleon $A$ and charge number $Z$. Furthermore, almost all massive stars are in binaries, or multiple systems \cite{Chini12,Chini13a,Chini13b}, while the binary frequency is reduced for lower mass stars. This naturally explains the rapid change in the chemical composition of CRs all across the knee \cite{CR-IV}, and a knee energy at about $10^{17.3 \pm 0.2} \, eV$ for Fe \cite{CR-IV,ASR18}; note that for the model worked out in 1993 the magnetic field in winds had not been known as well it is now. Today the magnetic fields are known to be stronger, and so all ensuing particle energies higher. Of those SNe that explode into the ISM, there are SN Ia that are exploding white dwarfs, and massive star SNe, that make neutron stars producing much lower particle energies. In the model of \cite{Gaisser13} there is a Galactic component of near EeV protons, that matches the model in \cite{CR-IV} using the better magnetic field numbers now known. A test has been made of this model in \cite{Thoudam16}. \cite{Allen24} focusses on those SNe, that explode into fast winds: On this basis we discussed there the new highly accurate AMS data on CRs. The sum of the CRS arising from ISM-SNe and wind-SNe, and their secondaries, gives structure to the spectrum at low energies \cite{CR-IV,Gal19,Allen24}. The essential message \cite{Allen24} is that almost all CR elements contain spallation secondaries, and we identified a spallation sequence, from a small secondary component, like for CR O, to a dominant secondary component like CR $^3$He. There is a large secondary component in CR protons.

The area around a rotating black hole (BH) has been observed by the \cite{EHT19L1,EHT19L5,EHT21a,EHT21b} and is found to be highly variable; in such a zone one may expect a population of energetic particles driven by stochastic processes, such as the second order Fermi process \cite{Fermi49,Fermi54}, followed by reconnection and other mechanisms (e.g., \cite{Meli08,MeliN21,Meli+23}). The particle energy may go up to the maximum which space allows for the Larmor motion. In Radio Super-Novae (RSNe) a wind of typically $10^{-5} \, {\rm M_{\odot} \, yr^{-1}}$ (a summary in \cite{ASR18}), a shock speed of about $0.1 \, c$, and a magnetic field of $(B \, \times \, r) \, = \, 10^{16.0 \pm 0.12} \, {\rm Gauss \, \times \, cm}$  are implied by the radio observations \cite{KBS85,Allen98,Allen99,Gal19}; these are the same numbers for Blue Super Giant star explosions as for Red Super Giant star explosions, despite their very different wind properties \cite{KBS85,Allen98,Allen99,Kronberg00,ASR18,Gal19}, as mentioned above. It is important to note that this latter quantity is independent of radial distance $r$ \cite{Parker58,Weber67}. In fact, the EHT data for M87 (\cite{EHT19L5}) as well the low power radio galaxies \cite{Punsly11} all show consistency with just this value of the quantity $(B \, \times \, r )$.   The quantity $(B \, \times \, r )$ is key for angular momentum and energy transport, and can be derived from the radio jet data. We note that just recently the super-massive black hole in M87 experienced a merger with another black hole, with a spin-flip visible in the data \cite{Owen00}; it might be possible that most, if not all radio galaxies evolve via many mergers of their central black holes as well as their host galaxies \cite{Rottmann,Gopal03,Gopal12,Jaroschewski23}. In the commonly used green onion model, in which a $2 \, \pi$ flow changes over to a jet flow (on both sides) we interpret the observations of RSNe  as a wind emanating from the black hole/accretion disk system and its near surroundings, after a stellar explosion which produced a rapidly rotating black hole (BH) \cite{Chieffi13,Limongi18,Limongi20}.  

The goal of this paper (see Table 1 for the run of the argument) is to understand the origin and the consequences of the quantity $(B \, \times \, r )$ showing the same number for stellar mass BHs and super-massive BHs, when we have reason to assume that in these cases the BH is rotating near maximum. Can we learn something about BHs from CR observations, and the answer we propose is "yes". We will propose an origin of the numerical value of $(B \, \times \, r )$ as rooted in a property of rotating BHs.


%
%
%
%


\begin{table}[h!]
\begin{center}
\caption{Run of arguments in this paper}
\begin{tabular}{|c|c|c|c|c|c|}
\hline
RSNe	&   show   &   $(B \times r)$&  = &  $10^{16.0 \pm 0.12}\, {\rm Gauss \times cm}$  \\
M87 BH	&   		&   the same & numbers  &     \\
Radio galaxies		&   low power & with & same & numbers        \\
\hline
	SMBH spin	&   radio galaxies&  show &  high  & BH spin     \\
\hline
Stellar evolution		& simulations  & yield  &  high &  BH  spin  \\
\hline
Proposal		&  in all  &  cases &  RSN & central   BH  \\
		& high spin  &  initially &   &     \\
\hline
Electric currents		& in  &  winds &  and  &jets     \\
		& driven  & by  drift&  $E^{-2}$ spectrum & of     \\
		& protons   & and  & anti-protons  &     \\
\hline
General Relativity		&  using radio  & observations  &   &     \\
		& using $(B \times r)$  &  constant with $r$ &  same & for all  \\
		& gives  & divergence  & of charged &  particle   \\
		&  density & near horizon &   &        \\
\hline
Collisions		& of protons  with & protons &  gives  & anti-protons       \\
		&  spallation of & heavier nuclei & so  & their destruction       \\
\hline
Prediction		& $E^{-2}$  &  source protons &  and  &  anti-protons   \\
		&   & $E^{-7/3}$  &  for observer &     \\
		&   &  reaching & EeV & energies    \\
		& identified  & from GeV  & to   & EeV  \\
\hline
\end{tabular}
\label{tablearguments}
\end{center}
\end{table}

\subsection{Black Holes}

A better understanding of the nature of black holes (BHs) has been sought ever since Schwarzschild's discovery \cite{Schwarzschild16} of a solution of Einstein's  equations \cite{Einstein15} with an essential singularity (black hole), and Kerr's generalization of the solution to a rotating BH \cite{Kerr63,Rees74,Rueda22}.  The most significant flaw is the failure to merge gravitational physics with quantum physics, with some convincing first steps \cite{Penrose71,Bekenstein73,Bardeen72,Hawking74,Hawking75,Rueda20,Rueda21}; some early and recent books are \cite{Misner73,Rees74,Joshi93,Joshi07,Joshi11,Joshi14,Joshi15}. The best hope to explore BH physics is to consider more detailed observations, e.g. \cite{Mirabel11}. The goal of this paper is to further the understanding of BHs by exploring the observations of Super-Nova Remnants (SNRs) which are produced in those SN explosions which lead to BHs. These sources are referred to as Radio Super-Novae (RSNe). Numerous observational data have been obtained for these stellar explosions, which make BHs, at various wavelengths.

There are a number of samples of Radio Super Novae (RSNe): First is the large set of RSNe in the discovery paper \cite{KBS85}, 28 sources certain, and 43 possible. Then there are the independent observations by the team of Muxlow, \cite{Muxlow94,McDonald02,Muxlow05,Muxlow10}, of the same population of RSNe in M82 (30 classified as SNR). There is the newly observed list of the M82 RSNe collected and analyzed in \cite{Allen98}. Then there are the lists assembled in \cite{ASR18} of Red Super Giant (RSG) and Blue Super Giant (BSG) RSNe, all from the literature. For the RSNe collected in \cite{ASR18} we know the moment of explosion, with all accompanying information; the radio interferometric observations (VLBI) have followed the expansion to a radial scale of order $10^{16} \, {\rm cm}$. For the RSNe in M82 we have only estimates when the explosion occurred, but from ram pressure arguments (\cite{Gal19}) one can show that these RSNe must have originated in most cases from the explosion of a Blue Super Giant (BSG) star; in these cases the expansion can be followed to a radial scale of order $10^{18.5} \, {\rm cm}$. All these RSNe are consistent with just different stages of the same kind of explosions, from RSG as well as BSG stars, for the large radial scales mostly BSG star explosions. Considering the independent data shown in the papers by Muxlow and his group (e.g. \cite{Muxlow94,McDonald02,Muxlow05,Muxlow10}), they give the same value of $(B \, \times \, r)$, just with a larger error bar, as the data obtained and analyzed by \cite{Allen98}.  The collection in \cite{Gal19} is based on the \cite{Allen98} analysis and data. Moreover, there is an independent discussion using another data set of very energetic explosions by \cite{Soderberg10b}, leading to about the same value for $(B \, \times \, r)$, as shown in \cite{ASR18}. 

It has been argued that very massive star SN lead to a BH by direct collapse, without leaving a visible trace (e.g., \cite{Smartt09,Smartt15,VanDyk17,Humphreys20}. These arguments are based on visual and infrared data, and are influenced by obscuration and selection effects. However, gamma-ray line data and radio data (e.g., \cite{Diehl06,Diehl10,Diehl11,Prantzos11,Diehl13,Siegert16b,ASR18}) clearly give much more accurate SN statistics data, unaffected by obscuration. These data show for instance (summarized in \cite{ASR18}), that Blue Super Giant star explosions happen in our Galaxy about once every 600 years, and in other galaxies at corresponding frequencies, scaled with the star formation rate, derivable from both far-infrared and radio observations, as they scale with each other (e.g., \cite{Tabatabaei17}).

These RSN range from RSG star explosions to BSG star explosions, which cover vastly different environments in density. Among the BSG star explosions they probably cover the entire range of masses (summarized in \cite{ASR18} based on the work of \cite{Chieffi13,Limongi18,Limongi20}), which can be derived from the now many lists in \cite{GWTC1,GWTC2,GWTC2p1,GWTC3}. Of course, the lists of observed mergers of stellar mass BHs encompasses second generation mergers, and that is why the BH mass can reach relatively high values, up to four times the highest single BH mass.

The only common feature of all these explosions is that they form a BH, and the explosions happen into a wind. SN-explosions that make a neutron star explode into the ISM. Here we consider explosions into a wind: and yet, the quantity $(B \, \times \, r)$ is consistent with having the same value for all explosions. We note that the EHT data for M87 are consistent with the same number; the radio galaxy M87 harbors a central black hole with a mass approaching $10^{10} \, M_{\odot}$, suspected to be near maximal rotation \cite{Daly19,EHT19L5}. The minimum jet powers in \cite{Punsly11} are also consistent with the same values.  So, we explore the possibility that this quantity is actually related to the BH in the sense that this quantity refers to a near maximal rotation of the black hole, independent of the mass, but with energetically negligible accretion. 

In the following we will assume that the physics around black holes scales such that fundamental principles carry over across all masses observed \cite{Merloni03,Merloni06,Falcke04,Markoff15,Gueltekin19}; this is commonly referred to as the ``Fundamental plane of black hole accretion". Much of the accretion physics is mass-invariant. As a consequence we will assume the same physical concepts across all masses of black holes discussed in the following.

\section{Radio Super Novae (RSNe) with freshly formed black holes (BHs)}

Where do we witness the formation of black holes? In massive star Super Novae (SNe), from stars of an initial mass (Zero Age Main Sequence or ZAMS) above about $25 \, M_{\odot}$ (at Solar abundances: \cite{Woosley02,Heger03,Chieffi13,Limongi18,Limongi20}), best observable as Radio Super Novae (RSNe). The radio data can be interpreted as follows: We observe a Parker wind, as $(B \, \times  \, r)$ follows two rules i) $(B \, \, \times \, r) \, = \, const.$ for a given RSN, over a range in radius, and also ii) that value is the same for different RSNe, in different galaxies and for very different radii $r$ \cite{Parker58,Weber67,ASR18,Gal19}; the best data are obtained from the starburst galaxy M82 \cite{KBS85,Allen98,Allen99}; the M82 sample can be checked also independently using the observations of \cite{Muxlow05}, and in other galaxies \cite{ASR18}; the radial range over which $(B \, \times \, r) \, = \, const.$ covers a factor of over 100.  The quantity is $(B \, \times \, r) \, = \, 10^{16 \pm 0.12} \, {\rm Gauss \, \times \, cm}$ \cite{Allen98,Allen99,Gal19}.  Probably all the Radio-Super-Novae (RSNe) detected in M82 can be traced back to BSG stars, all of which make black holes. This argument is based on the wind ram pressure, which is very much larger for a BSG star than for a Red Super Giant (RSG) star \cite{Gal19}. A wind from a RSG star is not expected to reach such large radii as parsec scale in an environment at a pressure like in the starburst galaxy M82. An expansion as in ISM-Super Novae \cite{Cox72} (i.e., SN exploding into the Interstellar Medium (ISM), the most common SNe) would not allow the quantity $(B \, \times \, r)$ to be constant; various other proposed explosion scenarios have been worked through in \cite{Gal19}; none of them allow to understand such a constant value for $(B \, \times \, r)$, independent of environment and of radius $r$. Furthermore, since the value of $(B \, \times \, r)$ is the same in all examples, in different locations in M82 as well as in different galaxies, also at a much earlier stage of RSN evolution, it is clear that the environment does not play a role in the expansion. The concept of a wind driven by a rotating compact object at its center \cite{Parker58,Weber67} has been generalized (e.g. \cite{Chevalier84}), to neutron stars \cite{Goldreich69}, to black holes \cite{BZ77} and to entire galaxies (e.g. \cite{Breitschwerdt91}). We note that the generic approach developed by \cite{Pacini73} in their development phase 2 gives a relationship as shown by the observations here, $( B \, \times \, r) \, = \, const.$, with the difference that the magnetic field is too high by an order of magnitude; however, the approach of \cite{Pacini73} was proposed for neutron stars which would be expected to yield somewhat different numbers as compared to BHs. \cite{Weiler80} applied this approach to the observations of supernova remnants driven by the slowing down of a central neutron star, which they called ``plerions".   Latest simulations are, e.g., those of \cite{Gammie20a,Gammie20b,Gammie21a,Gammie21b,Lucchini22,Cho23}. Much of this work focusses on the Magnetically Arrested Disk (MAD) models \cite{Igumenshchev03,Narayan03}, also postulating that the driver of activity is the spin-down of the central black hole. Here we focus on what the observations tell us about a wind driven by the central object in RSNe, a rotating black hole. The well established idea of a central spinning object driving activity by spin-down starting with \cite{Parker58} is used here as well. The key difference here is the observation that the magnetic field in terms of $(B \, \times \, r)$ appears to be the same value for the RSNe observed. 

In support of arguing that these RSNe contain BHs rotating near maximum, we note, that in radio galaxies it has been shown that the central BHs do rotate near maximum \cite{Daly19,EHT19L5}, with the same magnetic field directly measured or the magnetic field inferred in terms of the quantity $(B \times \, r)$ \cite{Punsly11} from the jet power \cite{Falcke95,Falcke04}.

We wish to emphasize here that all these RSNe clearly derive from a spectrum of BH masses, as the black hole merger data as well as the optical stellar observations of original stars show \cite{GWTC1,GWTC2,GWTC2p1,GWTC3,Chini12,Chini13a,Chini13b}. So the quantity $(B \, \times \, r)$ does not depend on the BH mass at its center. Massive stars producing black holes almost all start in a binary, triple or quadruple system, allowing the final BH initially near maximum spin from a tidal lock in the tight binaries \cite{Chini12, Chini13a,Chini13b,Limongi18,Limongi20}. Simulations suggest \cite{Limongi18,Limongi20} that the black holes formed may reach a high rotation rate, possibly  even slightly exceeding maximal just before a black hole is actually formed. 

\begin{figure}[htpb]
\centering
\includegraphics[scale=0.42]{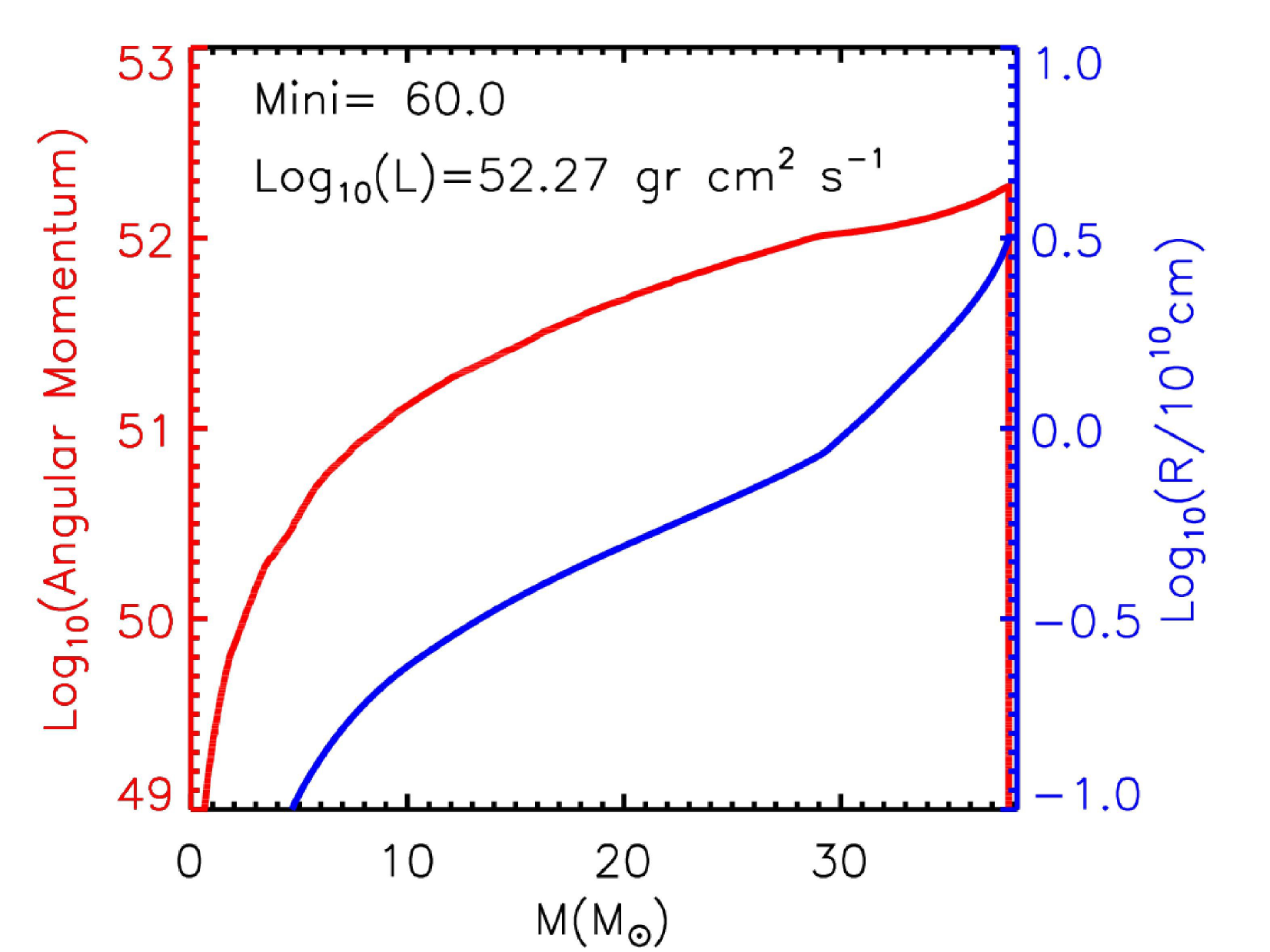}
\caption{Internal structure of 60 $M_{\odot}$ star just before making a black hole of 38 $M_{\odot}$. Source: Chieffi 2019 priv.comm., \cite{Limongi18,Limongi20}. {  Spin is $10^{52.27} \, {\rm erg \, s}$, a factor of $\sim \, 10^{0.21}$ over limit at $38 \, M_{\odot}$; relatively similar excess for other masses. Considering different radii each time the angular momentum is close to the maximum allowed for the mass contained in this radius; that means we have maximal differential rotation.}}
\end{figure}
\label{BlackHoleSpins}

\section{The EeV Cosmic Ray proton component}

At solar chemical abundances, stars $\simge \, 25 \, M_{\odot}$ Zero Age Main Sequence (ZAMS) mass evolve to RSG stars, while those $\simge \, 33 \, M_{\odot}$ ZAMS mass become BSG stars. Both classes of stars produce BHs \cite{Limongi18,Limongi20}. Stars between a ZAMS mass of about $\sim \, 10 \, M_{\odot}$ and $\sim \, 25 \, M_{\odot}$ produce neutron stars.

Magnetic fields \cite{Kronberg94,Kronberg16} are observed in the winds of massive stars (e.g., \cite{Maheswaran92}). Detailed further observation reveal, that massive stars are usually combined in binaries, triplets or quadruplets. This implies that these stars may lose orbital angular momentum efficiently, driving them progressively together - see the work in the group of Chini  \cite{Chini12,Chini13a,Chini13b,Barr13,Chini19}. Tidal locking then ensures that their rotation increases, resulting in the high rotation rates used in the simulations of Limongi \& Chieffi \cite{Limongi18,Limongi20}. These simulations show that massive stars can eventually lead to BHs which initially rotate near the maximum allowed \cite{Chieffi13,Limongi18,Limongi20}. 

We interpret the observed radio emissions as a wind, which is driven by a BH rotating near the maximum allowed via the Penrose/Blandford/Znajek mechanisms \cite{Penrose71,BZ77}.  This wind is thought of as keeping the energy and angular momentum transport processes functioning in the Radio Super-Novae (RSNe). The RSN data show that the slowest angular momentum transport time scale, derived from the afore-mentioned quantity $(B \, \times \, r) \, = \, 10^{16.0 \pm 0.12} \, {\rm Gauss \, \times \, cm}$, is $\sim \, 10^{3.7} \, {\rm yrs} \, (M_{BH}/M_{\odot})$, following \cite{Parker58} and \cite{Weber67}. Here we interpret the magnetic field observed, with $(B \, \times \, r) $ a constant without any indication of the magnetic field's direction, as $B_{\phi}$.  The specific number for the angular momentum transport time-scale depends on three factors, which together amount to a term between unity and ten in the case of near maximal allowed rotation. As a compromise number here we adopt the value of 5, with a large uncertainty. From the connection of mass, irreducible mass, and spin, we can derive in the limit of near-maximal rotation that $(d \, M_{BH})/(M_{BH} \; d \, t) \, =  \, (1/2) \, (d \, J_{BH})/(J_{BH} \; d \, t)$, where $J_{BH}$ is the angular momentum of the black hole. This gives a luminosity of $\sim \, 10^{42.8} \, {\rm erg/s}$, independent of BH mass in the mass range considered here.  This corresponds to within a factor of unity to the Poynting flux energy flow (e.g. \cite{Nokhrina20}); such an interpretation suggests that the wind is split into a fast jet along the symmetry axis and a slower wind around it, i.e. over much of $4 \, \pi$ (see the General Relativity Magneto-Hydrodynamic - GRMHD -  simulations of \cite{Moscibrodzka16,Davelaar18,Porth19}). The Super-Nova Remnant (SNR) data of Cas A in X-rays are compatible with this possibility \cite{Hwang04}. 

This is fully consistent with the voltage drop expected near a black hole (\cite{Lovelace76,Kronberg11}). In \cite{Kronberg11} the voltage near a black hole was worked out, and inserting the observed numbers corresponds to $10^{18.9} \, {\rm eV}$ independent of black hole mass. This value is consistent with the observed magnetic field strength in terms of $(B \, \times \, r)$.

To summarize the concept used here: When a massive star explodes, it explodes into its magnetic wind, which has pushed out a substantial fraction of its Zero Age Main Sequence (ZAMS) mass already prior to the explosion. A magnetic wind emanating from a compact object, here thought to be a rotating black hole, enhances the energy and angular momentum transport processes and provides an outward pressure. Thus all the primary CRs are accelerated in the SN shock, but an additional weaker ``special" CR component is proposed to come from the environment of the compact object, which we identify as 3$^{*}$ of \cite{Gaisser13}; but also refer to \cite{Thoudam16}. We propose that this component is currently also visible in data near EeV (see \cite{Auger20a}). This is indeed a small proton component \cite{Gaisser13}, if we think of pop 3$^{*}$, in either Table 2 or Table 3 of  \cite{Gaisser13}.

Noting that massive stars explode as SNe in our Galaxy on average about every 75 years (summarized in \cite{ASR18}; the error on these numbers is $10^{\pm 0.11}$: see \cite{Diehl06,Diehl10}), and those leading to BHs every 400 years (i.e. both RSG and BSG stars); we can check the energy budget. Here we take the numbers of \cite{Gaisser13}, which indicate that one needs to account for $10^{41} \, {\rm erg/s}$ for CR production in our Galaxy. Following these authors, we adopt 10 percent of the kinetic energy as leading to CR production. This then suggests that every massive star which makes a neutron star produces $10^{50}$ erg in CRs, i.e. $\sim \, 5 \cdot 10^{40} \, {\rm erg/s}$. Those stars which make BHs are then required to give  $10^{50.8} \, {\rm erg}$ in CRs of  to match the energy budget given by  \cite{Gaisser13}. This is in accordance with numerous observations of massive star SN explosions (see, e.g.,  \cite{Pavalas01}), in that they produce about an order of magnitude more energy than the more common SNe which lead to neutron stars (\cite{ASR18}). The CR production of SN explosions of type Ia has been discussed in \cite{Gal19}.

In this work, we will use the \cite{Gaisser13} tabular fits (their table 3). In \cite{Gaisser13} there is a population of protons, referred to as either "Pop. 3", with a cutoff energy of 1.3 EeV, a differential spectral index of 2.4, and a relative abundance of 0.002, or as "Pop. 3$^{*}$" with 1.5 EeV, 2.4, and 0.0017, respectively. For reference, we note that assuming a slightly flatter spectral index of $7/3 \, \simeq \, 2.33$ lowers these relative abundances by a factor of about 4, i.e. to $ 0.002/4 \, = \, 0.0005$ or $0.0017/4 \, = \, 0.000425$.  

The RSNe expand to about 1 to 2 pc \cite{KBS85,Allen98,Allen99}  with an observed shock speed of $c/10$ \cite{ASR18}, giving a time scale of $10^9$ to $10^{9.3} \, {\rm s}$, which in turn gives a total electromagnetic energy output of $10^{51.8}$ to $10^{52.1} \, {\rm erg}$, consistent with the numbers inferred above. This energy supply is similar to the SN mechanism of Bisnovatyi-Kogan \cite{GBK70}, which is worked out in, e.g., \cite{GBK08}, and many further papers. This is consistent with observations of the explosions of similar stars (in the starburst galaxy M82 Blue Super Giant stars, \cite{ASR18,Gal19}), mentioned above (see \cite{Pavalas01}, for an earlier demonstration of such energetics); these stars have a ZAMS (Zero Age Main Sequence) mass of $\simge \, 33 \, M_{\odot}$ at Solar chemical abundances \cite{Limongi18,Limongi20}. The observed RSN wind allows an energy flow of an energetic particle population of $\sim \, 10^{39.8} \, {\rm erg/s}$ at most, so the energy flow is down by $\sim \, 10^{-3}$ from the total energy flow. This corresponds to the CR population, "3" as well as "3$^{*}$" in  \cite{Gaisser13}. By fitting the Larmor motion diameter into the space available, we obtain a maximal energy of $(1/2) \, (e \, * B \, \times \, r) \, = \,  10^{18.15 \pm 0.12} \, {\rm  eV} $, which is the same quantity which rules angular momentum flow. This also matches the Gaisser fit to the maximum proton energy in the range of 1.3 to 1.5 EeV \cite{Gaisser13}.

These ideas are in good agreement with Auger.  The relevant statement  \cite{Auger20b} is that at energies below 1 EeV, even though the amplitudes are not significant, their phases determined in most of the bins are not far from the R.A. of the Galactic center - $RA_{GC} \, = \,  - \, 94 \, {\rm deg}$. This suggests a predominantly Galactic origin for anisotropies at these energies. The reconstructed dipole phases in energy bins above 4 EeV point instead to R.A.'s that are almost opposite to the Galactic center R.A.: They suggest a possible extragalactic CR origin  (cited nearly verbatim from \cite{Auger20b}). In the data analysis by  \cite{Gaisser13} the components "3" or "3*" referred to above have a cutoff at 1.3 to 1.5 EeV, and this is the component argued about here quite explicitly. Therefore this EeV CR proton component appears to be fully consistent with Auger data, and is in fact almost required by the data (see Fig. 5 in \cite{Gaisser13}). In a chemical composition analysis of the Auger data a proton component with such a cutoff is clearly detectable (see Fig. 2 in \cite{Auger20a}). The mixed chemical composition around the knee and above was predicted in \cite{CR-IV}, is consistent with \cite{Gaisser13}, confirmed in \cite{Thoudam16}, and is visible in the new Auger data \cite{Auger20a}.

\section{Why this value of ${ (B \, \times \, r)}$?}


\subsubsection{The magnetic field due to the convection}

The magnetic  field observed via non-thermal radio emission in the winds of massive stars (\cite{Abbott84,Drake87,Churchwell92}) can be attributed to the dynamo process working in the central convection zone of massive stars (\cite{CR-II}). The rotation and convection allows the magnetic field to be amplified right up to the stress limit. Then the magnetic field can meander in flux tubes through the radiative zone, and penetrate into the wind. The estimate gives the right order of magnitude, but does not allow to comprehend, that the resulting magnetic field observed in the post-shock region of the SN-explosion racing through the wind is the same number for very different stars, RSG and BSG stars, with extremely different wind properties.

\subsubsection{The magnetic field due to the SN-shock}

The magnetic field could be enhanced through the SN-shock itself, observed to be at a velocity of about 0.1 $c$ for both RSG and BSG star explosions \cite{ASR18}. The Bell-Lucek mechanism \cite{LucekBell00,Bell01} can certainly produce strong magnetic fields, but to give the same strength of the magnetic field in two very different types of winds is highly implausible; the ram pressure of the SN-shock in these two types of wind is orders of magnitude different due to the much higher density in RSG star winds than in BSG star winds, as they show about the same shock speed, and the same mass loss in the prior wind.

\subsubsection{The magnetic field due to the central object}

The central object and its immediate environment could also determine the magnetic field strength of the wind visible, just as in the Pacini \& Salvati \cite{Pacini73} approach. The observations show that all RSNe show the same magnetic field in terms of $(B \, \times \, r)$, a constant for $B_{\phi}$ throughout a Parker wind \cite{Parker58}, despite the fact that massive stars over a wide range of masses produce such SNe, including RSG stars with slow and dense winds \cite{ASR18}. Furthermore, the environment of the big black hole in the galaxy M87 also shows a magnetic field consistent with the same number in these terms \cite{EHT19L5}. This can speculatively attributed to the environment of a rapidly rotating black hole, rotating near maximum, and independent of the mass of the black hole. This magnetic field can be translated into a wind or jet power, and the magnetic field observed corresponds to the minimum jet power in radio galaxies \cite{Punsly11}. So it is plausible to interpret this number as due to a pure spin-down power, as done in \cite{EHT19L5}. This implies that radio galaxies relatively quickly revert to pure spin-down power after a merger of two central super-massive black holes, as demonstrated by the X-shape of the radio galaxy Cen A \cite{Gergely09,Gopal03}.

\subsection{Some important questions}

At this point there are some important questions: 

$ {\boldmath \bf \; 1:}$ What is the reason for the observed specific number $(B \, \times \, r) \, = \, 10^{16.0 \pm 0.12}$ $ {\rm Gauss \,\times \, cm}$? It can be written as an energy flow with $ (B \, \times \, r)^2 \, c \, = \, \{\hbar \, c\}/e^2 \,  \, \{m_X \, c^2 \}/ \tau_{Pl} $ with $m_X$ close to the proton or neutron mass, and $\tau_{Pl}$ the Planck time (see, e.g., \cite{Rueda21}).  Below, in the paragraph headed by "Frequency of the Penrose process" we will derive such a relationship based on angular momentum flow; this relationship supported by observations requires the Planck time, and so connects gravitation and quantum mechanics.

$  {\bf \; 2:}$ Is there a possible physical connection to a relationship between magnetic field and rotational frequency (here equivalent to radius at maximum spin) also well known for super-conducting spheres \cite{Hirsch14,Hirsch19}?  

$  {\bf \; 3:}$ Does this also explain that knee and ankle energy are independent of the mass of the star which explodes and makes a BH?  This has in fact been proposed (e.g. \cite{CR-I,CR-II,CR-III,CR-IV,ASR18,Gal19}). Finding the relationship between magnetic field and angular momentum transport, as explained here in this paper, provides this argument.

$  {\bf \; 4:}$ Do all BHs near maximal rotation have the same magnetic field in terms of $(B \, \times \, r) $ independent of mass?  That does seem to be the case, comparing magnetic field strength numbers in RSNe and in M87 (\cite{EHT19L5})) and the inferred energy flow in radio quasars (Punsly \& Zhang 2011). The relationship derived below supports this conclusion.

$  {\bf \; 5:}$ What is the magnetic field at lower spin?  Here the Galactic Center SMBH will be a useful test. This will be derived in a subsequent paper. Some dependencies on spin are derived below.

$  {\bf \; 6:}$ What is the effect of electric drift currents (\cite{Northrop63}, eq. 1.79) allowed by an energetic population of $E^{-2}$ particles? Such electric drift currents can occur in electrically neutral plasmas and can be extremely fast.  This was worked out in \cite{Gopal24}, where it was shown that electric gradient drift currents, electric fields, and violent discharges are quite common in variable jets and winds.

All this provides motivation for deeper study.

\section{Angular momentum transport}

Since the quantity $(B \, \times \, r)$ is strongly connected to angular momentum transport, we consider this next.

\subsection{A Parker limit approximation}

At first we consider a Parker limit approximation to understand what is required at the inner boundary even in the simple Newtonian limit approximation. In this case we can include the $\phi$-dependence, which we cannot do in the GR approximation. We posit

\begin{equation}
r^2 \, B_{r} \, = \, B_0 \, r_H^2 \, H(r - r_H) \, \{\cos \theta\} \, \{\cos \phi\}
\end{equation}

\begin{equation}
r \, B_{\theta} \, = \, - \, B_0 \, r_H^2 \, \delta(r - r_H) \, \frac{\sin \theta}{2} \, \{\cos \phi\} \, + \, B_1 \, r_H \, H(r - r_H) \, \frac{\sin \theta}{2} \, \{\sin \phi \}
\end{equation}

\begin{equation}
r \, B_{\phi} \, = \, B_1 \, r_H \, H(r - r_H) \, \{\sin \theta\} \, \{\cos \theta\} \, \{\cos \phi\}
\end{equation}

\noindent $r_H$ is the radius of the horizon, assumed at first to be independent of $\theta$. $H(r - r_H)$ is the Heaviside function, and its derivative is the $\delta$-function $\delta(r - r_H)$.  This allows the angular momentum transport $B_{\phi} \, B_r \, r^3$ to be of the same sign everywhere.  This construction immediately allows the divergence equation to be satisfied, and avoids any requirement for a monopole. In this solution the magnetic field stops at $r_H$, and does not penetrate inside. It is obvious that the magnetic field could be expanded into a long series, just as in Parker (1958), but these are simple first terms. This results in

\begin{eqnarray}
& & B_1 \, r_H \, H(r - r_H)  \left[2 \, \cos^2 \theta - \sin^2 \theta \right] \, \{ \cos \phi\} \, - \, B_1 \, r_H \, H(r - r_H) \, \frac{ \{\cos \phi\}}{2} \, \nonumber \cr &-& \, B_0 \, r_H^2\, \delta(r - r_H) \frac{\{\sin \phi\}}{2} \, = \, \frac{4 \, \pi \, r^2 }{c} \, j_r
\end{eqnarray}

This allows the surface integral of the radial current $j_r$ to be zero, separately in $\theta$ and $\phi$. It also shows that the electric current runs in the same direction, both at $\theta \, = \, 0$ and at $\theta \, = \, \pi$, both either outwards or inwards. The current scales with $+ \, B_1$ near the two poles, and is negative with $- \, B_1$ at the equator, with negative values in a broad equatorial band.

\begin{equation}
 - \, B_1 \, r_H \, \delta(r - r_H) \, \sin \theta \, \cos \theta  \, \{\cos \phi\}- \, B_0 \, {r_H}^2 \, \frac{H(r-r_H)}{r^2} \, \frac{\cos \theta}{\sin \theta} \, \{\sin \phi\} \, = \, \frac{4 \pi \, r}{c} \, j_{\theta}
\end{equation}

and

\begin{equation}
\left[ B_0 \, r_H^2 \, \frac{\delta(r - r_H)}{r-r_H} \, \frac{1}{2} \, + \, B_0 \, r_H^2 \, \frac{H(r - r_H)}{r^2} \right] \,  \sin \theta \, \{\cos \phi\} \, + \,  B_1 \, r_H \, \delta(r - r_H) \frac{\sin \theta \, \{\sin \phi\}}{2} \, = \, \frac{4 \, \pi \, r}{c} \, j_{\phi}
\end{equation}

Considering the $\delta$-function as a narrow Gaussian this suggests a double-layer in the $\phi$-current, plus an asymmetric term.

This clearly shows that already in this simple approximation we get a $\delta$-function term, and even the derivative of a $\delta$-function term for the electric current at the inner boundary. It also demonstrates that the density of the current carrying charged particles diverges at the boundary, which implies that collisions also diverge in this approximation. One part of the end-product of these collisions is accreted to the BH, and the other part is ejected in the wind with a known magnetic power flow independent of BH mass.

\subsection{A General Relativity solution}

Here we derive the angular momentum transport in the terms of General Relativity, so allowing to treat the behavior of the magnetic field close to the black hole, for any rotation. In this section we set the speed of light $c$ to unity for simplicity

The metric tensor elements for the Kerr metric are given in Boyer - Lindquist coordinates by 

\be 
{ds^2}&=&\frac{{d\phi }^2 \sin^2(\theta ) \left(\left(a^2+r^2\right)^2-a^2 \sin^2(\theta ) \Delta(r)\right)}{\rho(r,\theta )^2}-\frac{({dt} {d\phi }+{dt} {d\phi }) \left(2 a G_N M_{BH} r \sin ^2(\theta )\right)}{\rho (r,\theta )^2}\nonumber\\
&+&{d\theta }^2 \rho(r,\theta )^2+\frac{{dr}^2 \rho(r,\theta )^2}{\Delta (r)}+{dt}^2 \left(-\left(1-\frac{2\,G_N\,M_{BH} r}{\rho (r,\theta )^2}\right)\right)\, . \nonumber
\ee
where $G_N$ is the universal gravitational constant, $M_{BH}$ is the mass of the black hole and
\be 
\rho(r,\theta)^2\,=\, r^2+ a^2\,\cos^2(\theta)\, ,\nonumber\\
\Delta(r)\,=\, r^2 -2\, G_N\,M_{BH}\,r+a^2\, .\nonumber
\ee

The electromagnetic tensor is 
\be 
F_{\mu \nu}\, =\,\left( \begin{array}{c c c c} 
 0 & 0 & \tilde{E}_{\theta}(r,\theta)& 0 \\
0 & 0 & \tilde{B}_{\phi}(r, \theta) & -\tilde{B}_{\theta}(r, \theta)\\
-\tilde{E}_{\theta}(r,\theta) & -\tilde{B}_{\phi}(r, \theta) & 0 & \tilde{B}_r(r, \theta) \\
0 & \tilde{B}_{\theta}(r, \theta) & -\tilde{B}_r(r, \theta) & 0\\
\end{array}\right) \, ,\nonumber
\ee
and the components of $F_{\mu \nu}$ are determined from the vector potential components $A_{\mu}$ 
\be 
F_{\mu \nu}\,=\,\partial_{\mu}\,\left(\sqrt{g_{\nu \nu}}\,A_{\nu}(r, \theta)\right) - \partial_{\nu}\,\left(\sqrt{g_{\mu \mu}}\,A_{\mu}(r,\theta)\right) \, . \nonumber
\ee
The measured components of the electric and magnetic fields are related to the tilde components in $F_{\mu \nu}$ by the relations
\be 
\tilde{E}_{\theta}(r,\theta)&=&E_{\theta}(r,\theta)\nonumber\\
\tilde{B}_{r}(r,\theta)&=&\sqrt{g_{\theta \theta}\,g_{\phi \phi}}\,B^r(r,\theta)\nonumber\\
\tilde{B}_{\theta}(r,\theta)&=& -\sqrt{g_{r r}\,g_{\phi \phi}}\,B^{\theta}(r, \theta)\nonumber\\
\tilde{B}_{\phi}(r,\theta)&=& \sqrt{g_{r r}\,g_{\theta \theta}}\,B^{\phi}(r, \theta)\nonumber
\ee
These expressions are based on the definitions of the electric and magnetic fields given in Komissarov \cite{Komissarov04}.  They have the asymptotic forms given in Weber and Davis \cite{Weber67} We are assuming that the $r$- and $\phi$-components of the electric field are zero.  The $E_{\theta}(r,\theta)$ component of the electric field can be determined for the case of a static magnetic field, $\partial\,\vec{B}/\partial\,t\,=\,0$, from the relation
\be 
\left( \nabla\times\,\vec{E}\right)_{\phi}\,=\,0\, .\nonumber
\ee
This relation requires that
\be 
\text{E}_{\theta}(\text{r},\theta) \,=\, \frac{\text{E}_{0}}{\rho(\text{r},\theta)} \, ,\nonumber
\ee
where $\text{E}_{0}$ is a constant.  The $B^r(r,\theta)$ component of the magnetic field  is obtained from the divergence relation
\be 
\nabla\,\cdot\,\vec{B}(r,\theta)\,=\,0\, .\nonumber
\ee
For $B^{\theta}(r,\theta)\,=\,0$ \cite{Weber67} this relation requires that
\be 
B^r(r,\theta)\,=\, \frac{\text{B}_{0}}{ \sqrt{g_{r r}\,g_{\theta \theta}\,g_{\phi \phi}}}\, ,\nonumber
\ee
where $\text{B}_0$ is a constant.  The remaining components of the magnetic field are undetermined.  Based on observational radio data extensively discussed in \cite{ASR18,Gal19}, we assume that 
\be 
 \sqrt{g_{r r}\,g_{\theta \theta}}\,B^{\phi}(r, \theta)\,=\, \text{constant}\,=\,\text{B}_{p0}\, .\nonumber
 \ee
\paragraph{}
Here both $B^{r}$ and $B^{\phi} \, \sim \, \Delta^{1/2}$.  The ratio between $B_{\phi}$ and $B_r$ is given by the Parker model, and this indicates that $B_{p 0}/B_{0} \, \sim \, \chi/M_{BH}$, where $\chi$ is the dimensionless spin (i.e. maximum unity), so $\chi \, = \, a/M_{BH}$. Furthermore we assume that the total radial magnetic field energy is proportional to the available rotational energy, which results in $B_0 \, \sim \, \chi \, M_{BH}$, in the $\chi \, << \, 1$ approximation. From this it follows that $B_{p 0} \, \sim \, \chi^2$. Using observations of radio loud quasars \cite{Punsly11} we can check on the implications, since GR solutions and far-distant solutions have to be consistent in their dependence on $\chi$ and $M_{BH}$, namely $L_{jet} \, \sim \, \chi^4$, independent of BH mass $M_{BH}$. Furthermore $E_{0} \, \sim \, \chi^2 \, M_{BH}$ from the consistency requirement of the energy flow and angular momentum flow, worked out below.

The energy flux is obtained from the contraction of the covariant form of the Killing vector $k^{\mu}_{t}$ with the electromagnetic energy-momentum tensor
\be 
{\mathcal{E}^{\mu}}\,=\,T^{\mu \nu}\,(k_{t})_{\nu}\, , \nonumber
\ee
and the angular momentum flux is obtained from the contraction of the covariant form of the Killing vector $k^{\mu}_{\phi}$ with the electromagnetic energy-momentum tensor

\be 
{\mathcal{L}^{\mu}}\,=\,T^{\mu \nu}\,(k_{\phi})_{\nu}\, . \nonumber
\ee
The $r$- and $\theta$-spatial components of the energy flux and the angular momentum flux are given by
\be  
{\mathcal{E}}^r&=&\frac{\text{B}_{p0} \,\text{E}_{0}\, \Delta (r)}{\rho (r,\theta )^5}\nonumber\\
{\mathcal{E}}^{\theta}&=&0\nonumber\\
{\mathcal{L}}^r&=&\frac{\text{B}_{0}\, \text{B}_{p0}\, \Delta (r)^{3/2}}{\rho (r,\theta )^5}\nonumber\\
{\mathcal{L}}^{\theta}&=&0 .\nonumber
\ee
The energy flux and the angular momentum flux are related via the expression
\be 
{\mathcal{E}}^r\,=\, \omega(r,\theta)\,{\mathcal{L}}^r\, ,\nonumber
\ee
where 
\be 
\omega &=&\frac{\tilde{E_{\theta}}}{\tilde{B_r}}\, \nonumber \\
 &=& \frac{\text{E}_{0}}{\text{B}_{0} \sqrt{\Delta}}\, .\nonumber
 \ee
 This is the same relation as the one in Eq.(4.4) of \cite{BZ77}. Here $\omega \, \sim \, \chi/M_{BH}$.
 \paragraph{}

\begin{figure}[ht]
\center
\includegraphics[height=12.0cm]{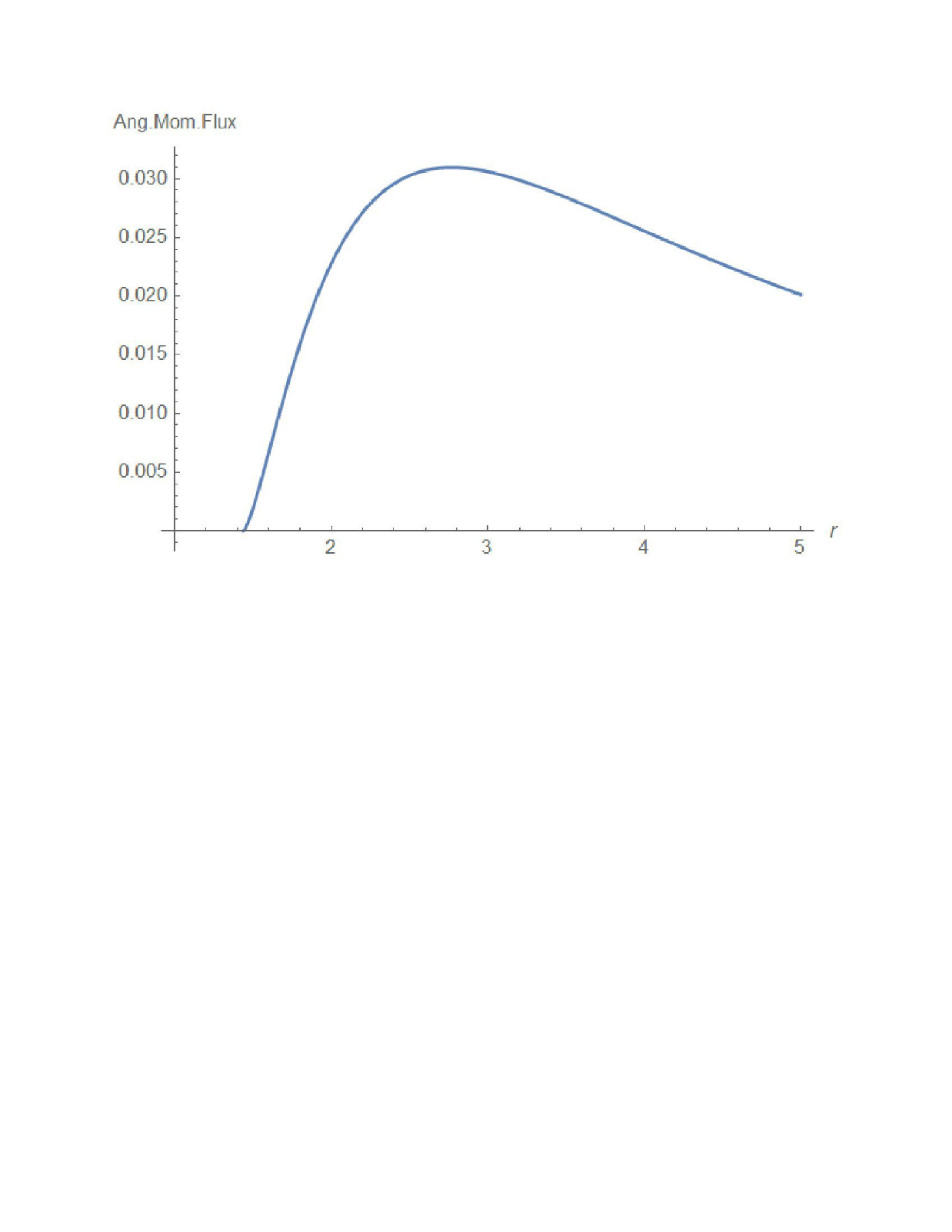}
\vspace{-5cm}

\caption{Radial component of the angular momentum flux vs the radius at the equator of the black hole ($\theta\,=\,\pi/2$).  In this specific plot as in the next two plots in the abscissa the unit is the Kerr radius, in contradiction to the text, where $r$ scales to the Kerr radius; so there $r$ has as a minimum the Kerr radius $\{G_N \, M_{BH}\}/c^2$, but with $c$ set to unity in this section. The ordinate is determined by the mathematical expression, setting all other constants to unity. The angular momentum per unit mass constant, $a$, is $a = 0.9$.    }\label{AMF}
\end{figure}

The location of the horizon is determined by the condition $\Delta(r)\,=\,0$, so the flux components components ${\mathcal{E}}^r$ and ${\mathcal{L}}^r$ vanish on the horizon.  On the equator of the black hole ($\theta\,=\,\pi/2$) the radial component of the angular momentum flux reaches a maximum at a radius of slightly less than three horizon radii, Fig.(\ref{AMF}).  These expressions are similar to the ones obtained by Blandford and Znajek \cite{BZ77}, but there are significant differences due to the differences between our model and theirs. In the BZ model both of the poloidal components of the energy flux are non-zero, while in our model both of the fluxes in the $\theta$-direction (polar direction) are zero. The vanishing of the $\theta$-component of the energy flux in our model is due to setting the $r$- and $\phi$-components of the electric field equal to zero, and the vanishing of the $\theta$-component of the angular momentum flux is due to setting the $\theta$-component of the magnetic field equal to zero, following Weber and Davis \cite{Weber67}. 
\vspace{1mm}

\subsection{Calculation of energy extraction and angular momentum extraction}

As seen by an observer at infinity the rate of energy extraction is given by
\be 
\dot{E}_{rad}\,=\,\int {\mathcal{E}^r \rho(r,\theta)^2\,d\Omega}\, , \nonumber
\ee
and the rate of angular momentum extraction is given by
\be 
\dot{L}_{rad}\,=\,\int {\mathcal{L}^r \rho(r,\theta)^2\,d\Omega}\, , \nonumber
\ee
where $d\Omega$ is the infinitesimal solid angle. The evaluation of these integrals gives (note that the radius $r$ refers to the BH mass, so that spin $a$, radius $r$, and $G_N M_{BH}$ have the same unit in this section)
\be 
\dot{E}_{rad}&=&\frac{4 \pi  \text{B}_{p0}\, \text{E}_{0}\, \left(a^2+r (r-2 G_N M_{BH})\right)}{r^2 \sqrt{a^2+r^2}}\nonumber\\
\dot{L}_{rad}&=&\frac{4 \pi  \text{B}_{0}\, \text{B}_{p0}\, \left(a^2+r (r-2 G_N M_{BH})\right)^{3/2}}{r^2 \sqrt{a^2+r^2}}\, .\nonumber
\ee

Here $\dot{E}_{rad} \, \sim \, \chi^4$, and $\dot{L}_{rad} \, \sim \, \chi^3 \, M_{BH}$, consistent with a derivation following \cite{Weber67} and \cite{Falcke95}.

\begin{figure}[ht]
\center
\includegraphics[height=14.2cm]{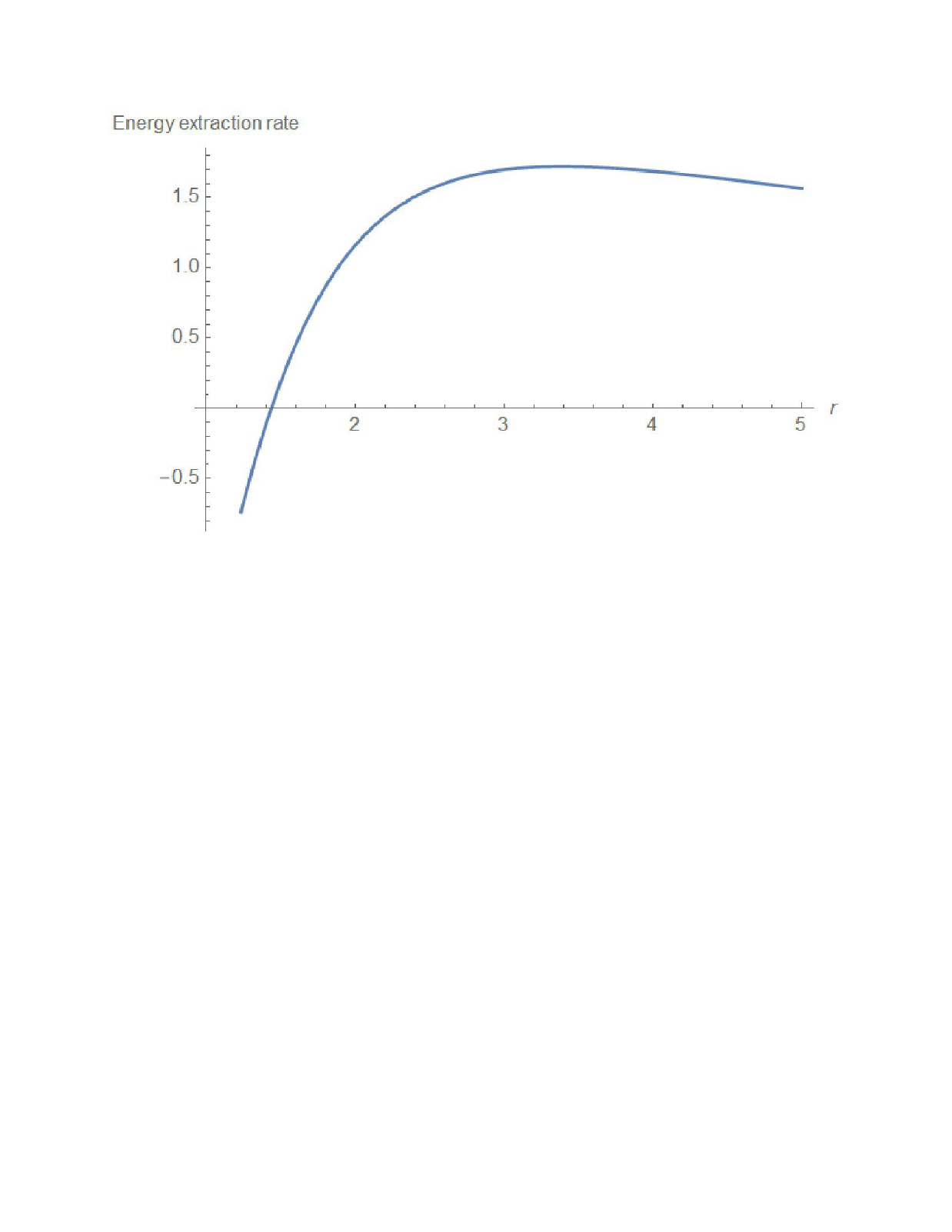}
\vspace{-8cm}

\caption{ Radial component of the magnitude of the rate of energy extraction.  The angular momentum per unit mass constant, $a$, is $a = 0.9$.  All other constants are set equal to 1.  The extrapolation to negative values of this extraction rate is without consequence for an observer, as this part of the curve is inside the horizon.  }\label{Enrext}
\end{figure}
\begin{figure}[ht]
\center
\includegraphics[height=14.2cm]{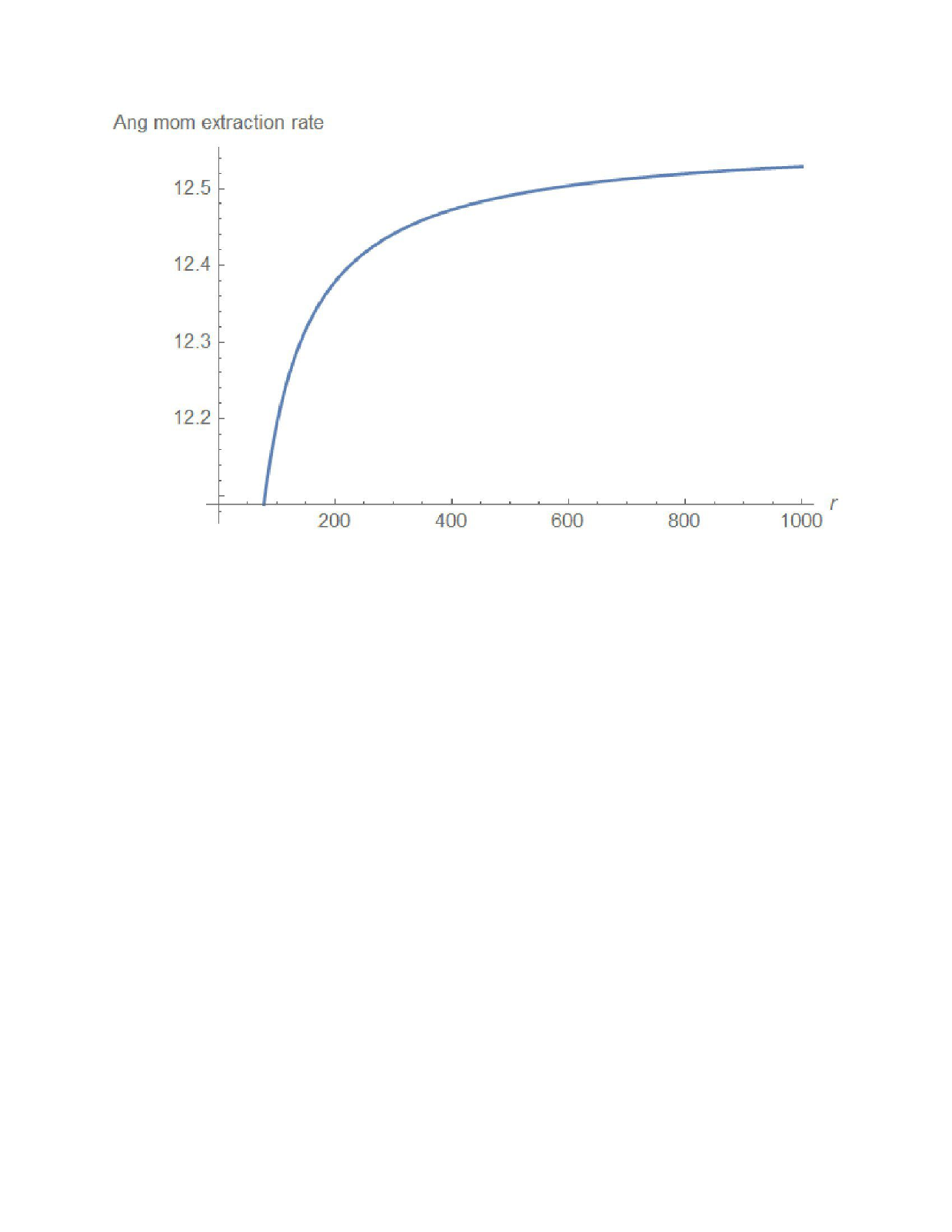}
\vspace{-8cm}

\caption{ Radial component of the magnitude of the angular momentum extraction rate vs the radius.  The angular momentum per unit mass constant, $a$, is $a = 0.9$.  All other constants are set equal to 1. Here the radial range considered is very large to show how this angular momentum transport approaches a constant despite the simplicity of the model.}\label{Angext}
\end{figure}
\vspace{1cm}

Here the power is proportional to $\chi^4$ and the angular momentum transport to $\chi^3 \, M_{BH}$. Since the power put out via magnetic fields is also proportional to $B_0^2$ \cite{Falcke95} this is consistent. The angular momentum transport by magnetic fields (\cite{Weber67}, eq 9, integrated over $4 \, \pi \, r^2$) runs as $B_0 \times B_{p 0} \sim \chi^3 \, M_{BH}$, so this is also consistent. In these graphs (Figs, 2, 3, 4) the lower limit of $r$ is given by the condition $\Delta(r) \, = \, 0$, so for maximal spin, that radius is $r \, = \, \{G_N \, M_{BH}\}/c^2$, the Kerr radius.

\subsection{Calculation of the current}
The current can be calculated from the covariant divergence of the electromagnetic field tensor
\be 
\nabla_{\mu} F^{\mu \nu}\,=\, J^{\nu} \nonumber
\ee

For the radial and theta components of the current this calculation gives
\be 
J^r&=& -\frac{4\, a^2\, \text{B}_{p0} \sin (\theta ) \cos (\theta ) \left(a^2+r\, (r-2\, G_N M_{BH})\right)}{\left(a^2 \cos ^2(\theta )+r^2\right)^3}\nonumber \\
J^{\theta}&=& -\frac{2\, \text{B}_{p0} \left(a^2 \cos ^2(\theta ) (G_N M_{BH} -r)+r \left(2\, a^2+r\, (r-3\, G_N M_{BH})\right)\right)}{\left(a^2 \cos ^2(\theta )+r^2\right)^3} \nonumber
\ee 
The $J^t$ and $J^{\phi}$ components are non-zero, but their expressions are much longer.  The latter two components decrease much more rapidly with $r$ than either $J^r$ or $J^{\theta}$. 

\subsection{Charge density}

The expression for the charge density as obtained from the covariant divergence relation is given by

\be 
& & J^0 = \nonumber \\
& &\frac{2 \sqrt(2)\,a^2\,\sin(2\,\theta)\,  (6 a^4\, \text{E}_{0}\, \cos ^2(\theta )-20\, a\, \text{B}_{0}\, G_N\, M_{BH}\, r\, \sqrt{a^2+r\, (r-2\, G_N\, M_{BH})}}{\left(a^2+r\, (r-2\, G_N\, M_{BH})\right) \left(a^2 \cos (2 \theta )+a^2+2\, r^2\right)^{7/2}}\\ & & +\frac{2\, \text{E}_{0}\, r^3\, (4\, G_N\, M_{BH} +3\, r))}{\left(a^2+r\, (r-2\, G_N\, M_{BH})\right) \left(a^2 \cos (2 \theta )+a^2+2\, r^2\right)^{7/2}}\nonumber \\
& &+\frac{2 \sqrt(2)\, a^2\, \sin(2\,\theta)\,(a^2\, \text{E}_{0}\, r)\, (14\, G_N\, M_{BH} + 9\, r + 3\, (-2\, G_N\, M_{BH} + r)\, \cos(2 \theta))}{\left(a^2+r\, (r-2\, G_N\, M_{BH})\right)\, \left(a^2\, \cos (2\, \theta )+a^2+2\, r^2\right)^{7/2}}\nonumber
\ee

This shows that in terms of the local charge density we also get a divergence at the inner boundary, at the horizon.  This is proportional to $\chi^4 \, M_{BH}^{-2}$. In more detail the leading terms with $B_0$ as well as $E_0$ run as $\chi^4 \, M_{BH}^{-2}$, while the terms with $E_0$ have two further terms running as $\chi^8 \, M_{BH}^{-2}$ and $\chi^6 \, M_{BH}^{-2}$.  It follows that the density may get high enough for lots of energetic collisions.

Furthermore the term running with $B_0$ has the factor $\Delta^{-1/2}$, while the terms running with $E_0$ all have the factor $\Delta^{-1}$. When $\Delta$ approaches a value small compared to radius $r$, and writing the spin parameter as $\chi \, = \, 1 \, - \, \delta \chi$ with $\delta \chi \, << \, 1$, then $\Delta \, = \, (r \, - \, r_g (1 \, + \, \sqrt{2 \, \delta \chi})) \times (r \, - \, r_g (1 \, - \, \sqrt{2 \, \delta \chi})) $. Writing the first term in brackets as $\delta r$, then $\Delta $ becomes $\delta r \times (\delta r + 2 \sqrt{2 \, \delta \chi})$. If we could constrain the collision rate then it follows that we could also constrain $\delta \chi$ to be a possibly small number.

To work out the numbers we note that $B_{p 0}$ is observed to be $10^{16.0 \pm 0.12} {\rm Gauss \times cm}$ \cite{ASR18,Gal19}; writing all other terms with their proper dimensions using the equatorial outer radius of the ergo-region of a $10 \, M_{\odot}$ BH, so $10^{6.4} \, {\rm cm}$, gives a charged particle density of about $10^{14.0} \, {\rm cm^{-3}}$, ignoring here the factors with some power of $\Delta$, and adopting the limit $\chi \, \simeq \, 1$.

This suggests that collisions could an important process, and this is what we explore further.

\subsection{Neutrinos from the ergo-region?}

There is an inconsistency between what the mass transport is in the wind (assuming equipartition with the observed magnetic fields) and what accretion to the BH is needed to sustain the luminosity of $\sim \, 10^{43.} \, {\rm erg/s}$, if one were to power this emission simply by accretion, as equality would require 100 \% efficiency. This inconsistency can be resolved by considering the pure spin-down mode \cite{BZ77}, which implies very little accretion. Here we note that in the pair production variant to the Penrose process, this could imply that the BH accretes predominantly particle/anti-particle pairs,  most of which never get out. The creation of such pairs costs at least two proton masses in energy, but energetically pion production dominates by far (below we use a factor of about 30 based on the ratio of cross sections to make pions and to make proton-anti-proton pairs from p-p collisions). They are available from interaction with magnetic irregularities and non-linear waves, such as shock waves. In fact, from the mismatch in mass turnover, one might speculate that the energetic protons initiate a cascade process similar to the interaction of 
ultra high energy CR particles entering the atmosphere of the Earth. In such a cascade a very large number of secondary particles is produced. By analogy with the Penrose argument one may expect that half the cascade particles are directly on orbits falling into the BH; the other half are initially on orbits to escape. These particles interact with the magnetic field. At the outer boundary of the ergo-region, the particles may transfer a significant fraction of their energy and angular momentum to the magnetic fields and fall back down in accretion to the BH (see \cite{Penrose71}).  In processes such as $\bar{{\rm p}}$'s colliding with ${\rm p}$'s, pions and multiple neutrinos are produced. These neutrinos have a good chance to escape altogether. All this should be re-evaluated using proper frames (e.g., \cite{Bardeen72,Shaymatov15,Bambhaniya21}), although a collision-dominated gas with a magnetic field, in which some energetic particles have Larmor radii which are close to the scale of the system, is a challenge.  What we present here is a detailed balancing of different particle species in the local frame.

{
Many different losses go into production of pions, which quickly decay into energetic electrons, positrons, photons and neutrinos. In the model proposed the photons are optically thick in their propagation. This is akin to the model published for blazars, and their neutrino emission in \cite{Kun21}.  The electrons/positrons and neutrinos have a chance of escaping. Based on the ratio of cross-sections for p-p-collisions to make pions versus p-p-collisions to make proton-anti-proton pairs, about 30 times as much energy goes into an electron/positrons pair plasma from pion decay (ratio of cross sections and energy turnover), and neutrinos, as goes into proton-anti-proton pairs, in terms of what gets out.  The neutrinos  - in the model proposed - range from MeV to very much higher energy, and for those the IceCube data provide a serious upper limit, if the model is used at TeV energies and beyond. Other than an electron/positron plasma neutrinos could be a second main escape path. That is a main point of the model.  

A check with data can be done: the proposal is consistent with IceCube  \cite{IceCube16,IceCube21} and INTEGRAL data \cite{Diehl06,Diehl10,Siegert16a,Siegert16b}:

In the model proposed the cosmic ray flux of the component going to EeV energies is about $10^{-2.8}$ of the normal CR flux at GeV energies (numbers taken from \cite{Gaisser13}, table 3, CR components 3 or 3$^{*}$; pop 3 contains all elements (Table 2) and pop 3$^{*}$ contains only protons); correcting for a slightly flatter spectrum assumed here, anchored at EeV, gives about $10^{-3.4}$. This implies $10^{37.6} \, {\rm erg/s}$, again using Gaisser's et al. numbers for the entire Galaxy of $10^{41} \, {\rm erg/s}$. \cite{Falcke13}  give an estimate of the accretion rate measured close to the central BH in our Galaxy, and it corresponds to a power of about $10^{37.8} \, {\rm erg/s}$, consistent with the number above.  INTEGRAL \cite{Siegert16b} gives a positronium production of $10^{43.5} \, {\rm s^{-1}}$ in a very large region, with a scale height of several kpc and along the plane from a larger region than any other recognizable source class, corresponding to about $10^{37.5} \, {\rm erg/s}$, again consistent with the \cite{Gaisser13} number. The papers by Diehl et al. support the point of view that there could be plenty more electrons and positrons that escape from the Galactic disk unseen. The production of a large number of electrons/positrons is demonstrated by observations of the BH V404 Cyg \cite{Siegert16a}. The electron/positron pair plasma production in our Galaxy appears to be due to many sources, possibly the Galactic Center black hole (GC BH) and most probably many stellar/SN/BH sources, including microquasars and SN Ia supernovae \cite{Martin10,Prantzos11,Prantzos17,MeraEvans22}. Diehl et al. propose that all black holes produce an electron/positron pair plasma, often in outbursts.  Based on gamma-ray line spectroscopy \cite{Diehl06,Diehl10,Diehl17} give a SN rate of those SNe making black holes in the Galaxy of about 1 SN per 400 years (again, with an uncertainty of $10^{\pm 0.11}$); this has been worked through in \cite{ASR18}; this includes both Red Super Giant and Blue Super Giant star progenitors, both of which produce black holes, or short BH-SNe. The time scale of the activity is at least 30 years (1 parsec at $0.1\, c$), as observed numbers from Radio Super-Novae given in \cite{Gal19}, based on the M82 data of Radio Super-Novae (RSNe) \cite{Allen98,KBS85,Kronberg00}. It ensues that each BH-SN contributes - again using the numbers in \cite{Gaisser13} - about $10^{50.8} \, {\rm erg}$ in CRs, as shown above. For this specific low level HE CR component this translates to $10^{47.4} \, {\rm erg}$, as well as $10^{48.9} \, {\rm erg}$ in e$^+$ e$^{-}$ plasma and MeV neutrinos, by virtue of the 30 times larger cross section (p-p collisions making pions versus p-p collisions making $p$-$\bar{p}$ pairs). This translates into a maximal flux, using the shortest reasonable time scale - of $10^{39.9} \, {\rm erg/s}$ initially. The observed power of about $10^{37.5} \, {\rm erg/s}$ in the Galactic Center region (by INTEGRAL) in electron/positron plasma means, if produced by a SN,  that the activity could be down now by $e^{-400/30}\, \simeq \, 10^{-5.8}$, for a possible initial power of $10^{43.3} \, {\rm erg/s}$ for all SN contributors summed together. This in fact approximately matches the spin-down power seen in both M87 \cite{EHT19L1,EHT19L5}, many other radio galaxies in their minimum jet power (e.g. \cite{Punsly11}),  and in Radio Super-Nova Remnants interpreting them as driven by a relativistic wind from a spinning compact object, presumably a BH. 


Using the starburst galaxy M82 (\cite{KBS85,Allen98,Kronberg00}) itself as our IceCube limit for point sources \cite{IceCube16,IceCube20,IceCube21} gives about $10^{-12.0}$ TeV events ${\rm cm^{-2} \, s^{-1}}$, assuming a $E^{-2}$ spectrum, corresponding to a limit of about $10^{-11.0} \, {\rm erg \, cm^{-2} \, s^{-1}}$ at GeV for a $E^{-7/3}$ spectrum assumed here for the relevant CR spectrum, where the \cite{Gaisser13} numbers are anchored. This corresponds to a limiting luminosity at TeV of $10^{39.1} \, {\rm erg/s}$ at the distance of M82, and $10^{33.9} \, {\rm erg/s}$ at the distance of the Galactic Center. Since there is evidence from the Telescope Array \cite{TA20} as well as Auger \cite{Auger18}, that both starburst galaxies M82 in the North and NGC253 in the South may have been detected in UHECRs, we assume that the detailed analysis of recent Radio Super-Novae (RSNe) in M82 applies also to NGC253 \cite{KBS85,Allen98,Kronberg00}, where the specific IceCube limit mentioned above applies  and so a limit for all sources is $< \, 10^{39.1} \, {\rm erg/s}$. In M82 there are about 40 such sources \cite{KBS85}, so the limit per source is $< \, 10^{37.5} \, {\rm erg/s}$, if all sources contribute equally. However, again, for a possible decay time of 30 years, only one source may contribute, and this possibility would imply a luminosity of $< \, 10^{39.1} \, {\rm erg/s}$ for that one source. To within the large errors of such an estimate this is still consistent with the data, which give an expectation for a single contributing source at $10^{39.9} \, {\rm erg/s}$. Allowing for a slightly steeper spectrum would loosen these constraints, as would an even faster change with time of any single source. The age of the youngest source 41.9+58 is sufficiently large so that it may have decayed already significantly.  Of course, if the HE neutrinos were pointed in their emission, then their luminosity could be quite a bit higher without showing up in our observations.

To do a further test: Applying the same neutrino flux limit to possible sources in the Galactic Center (GC) region gives a limit of about $10^{5.1}$ times stronger, so $< \, 10^{34.4} \, {\rm erg/s}$. As shown above this is fully consistent with the rate of BH-SNe occurring; the expected flux reduction is $10^{-5.8}$ for an initial luminosity limit of $< \, 10^{40.3} \, {\rm erg/s}$, again consistent. One problem in such an argument is that the sources are known to be highly fluctuating (e.g., \cite{Siegert16a}). It is possible to repeat this exercise for the Cyg region, which is much closer than the Galactic Center. This gives a limiting luminosity of $10^{33.2} \, {\rm erg/s}$, and it is again consistent, since the BH-SN rate is very low near to us, 1 BH-SN per about $10^5$ years, so predicting a huge reduction from the expected initial luminosity of $10^{39.9} \, {\rm erg/s}$ worked out above; Cygnus might be close enough to provide an actual source of the Galactic EeV CRs identified by \cite{Gaisser13}.

\subsection{Collisions}

Analyses of particle collisions near to BHs and singularities have been carried out, \cite{Patil10,Patil11a,Patil11b,Patil12a,Patil12b,Patil14,Patil15,Patil16,Liu11,Banados09,Banados11}.  These papers did not have the benefit of insight provided by the RSN observations, the most detailed of which by Allen \cite{Allen98,Allen99,Kronberg00}. The latter provide a newer solid foundation to develop the approach.

As an example, we calculate the particle density and flux for the ergo-region around a stellar mass BH of $10 \, M_{\odot}$: The magnetic field, extrapolated to near the BH, at radius $R \, = \, 10^{6.4} \, {\rm cm}$, is about $10^{9.6} \, {\rm Gauss}$. In equipartition, $(B^2)/(8 \, \pi) \, \simeq \, n \, k_B \, T$, leading to a particle density of $n \, \simeq \, 10^{21.6} \, {\rm cm^{-3}}$ at a weakly relativistic temperature of $\sim \, 10^{12} \, {\rm K}$. This, in turn, allows a flow of particles of $4 \, \pi \, R^2 \, \, n \, c \, \simeq \, 10^{46} \, {\rm s^{-1}}$. Interactions give a similar number, using a cross section of $10^{-27} \, {\rm cm^2}$ (valid for making proton-anti-proton pairs \cite{Winkler17,Reinert18}; the inelastic cross-section is about 30 times higher well above threshold), as obtained from $4 \, \pi \, R^3 \, n^2 \, \sigma \, c \, \simeq \, 10^{47} \, {\rm s^{-1}}$, which is more than what is needed to explain the observations; as even a smaller cross-section could be accommodated. This latter quantity cannot be readily extrapolated to a higher BH mass, as we discuss below.

Using the general approach of \cite{EHT19L5} we can show that this optical depth may reach order 10, independent of radius. This means that the interaction time to produce proton-anti-proton pairs is less than the residence time, possibly considerably less.

The observations show that $B \, = \, {10^{16.0 \pm 0.12}}/{r} \, {\rm Gauss}$.
\noindent with $r$ in ${\rm cm}$. This relationship has been observed over the range of radius from about $10^{18.5} \, {\rm cm}$ down to order $10^{16} \, {\rm cm}$, with the highest resolution observations done by radio interferometry (VLBI). Using the analogy with the Solar wind \cite{Parker58,Weber67} we extrapolate it down for the case of fast rotation. The EHT observations of M87 suggest that such an extrapolation is reasonable \cite{EHT19L5}: There the product $(B \, \times \, r)$ has about the same value as in RSNe at about 5 gravitational radii; the M87 black hole has been suspected to be in substantial rotation, perhaps near maximal (\cite{Daly19,EHT19L5} and later). The jet power of M87 is consistent with what is derived for RSNe using the available energy content of a maximally rotating black hole, and the time-scale derived from angular momentum transport (\cite{Weber67}). This suggests that the jet power far outside the ergo-region is already visible at five gravitational radii.

Putting in numbers as observed (\cite{EHT19L5}) extrapolated to a stellar mass BH  suggests that the production time scale for making proton-anti-proton pairs is safely of order  $ < \, 1$ of the resident time scale in the inner region around the ergo-region.

This argument works for stellar mass black holes, and we can speculate here that the model proposed would allow this to work also for more massive black holes.}

\subsection{Anti-protons}

The concept is that the energetic particles are confined by the magnetic field and so stay in the ergo-region; the magnetic field is due to electric currents in the (weakly relativistic) thermal matter, which is held in the gravitational field. In momentum phase space there is a cone, inside of which all particles are on orbit to accrete to the BH. This is akin to arguments in \cite{Hills75,Bahcall76,Frank76}. In that approach, stars interact with molecular clouds to fill a cone in momentum phase space which allows accretion to a central BH. This is referred to as the {\it loss cone} mechanism. Here, charged particles interact with the magnetic fields \cite{Strong07,Moskalenko19}, and also with each other, to also finally accrete to the BH.

Given all the above arguments, what are the predictions in these scenarios? In these conditions, one can ask what the fraction of anti-protons $n_{\rm \bar{p}}/n_{\rm p}$ might be. The observed fraction of anti-protons is about $10^{-3.7}$ \cite{AMS16}, with a spectral shape dependence of about $E^{-2.7}$ for both protons and anti-protons. We assume that this spectrum changes for both towards a flatter spectrum at higher energy since at lower energies, both components have other contributions (see, e.g., \cite{ASR18}). Could this match the observed flux of anti-protons? Fitting above 200 GeV, the CR flux is about $10^{-3}$ relative to other CR-populations from the similar SN-explosions.  Using a spectrum such as $E^{-7/3}$, this modifies the factor of $10^{-3}$ to $10^{-3.3}$ to $10^{-3.4}$. However, at EeV, the sum of protons and anti-protons is observed, while at lower energy, anti-protons are observed separately. Thus, correcting the prediction by another factor of order 2 gives $10^{-3.6}$ to $10^{-3.7}$, which allows the observed $10^{-3.7}$. Consequently, we propose a model to explain the flux, energy content, spectrum, maximal particle energy, and particle/anti-particle ratio of highly energetic protons. It follows then, that the spectrum of anti-protons continues all the way to ankle energies, with a spectral shape near $E^{-7/3}$. The energetic protons would approach the spectrum of the anti-protons at some energy slightly above PeV. AMS may well detect some of these anti-protons among its highest energy particles, around TeV.

One may well ask whether anti-protons survive their path to us: Their cross-section to interaction is the same as for protons, and since we see protons at EeV \cite{Auger20a} without being able to distinguish protons and anti-protons, the particles detected may well contain anti-protons, in this proposal here possibly half.

If there are in fact large numbers of cascades, then many of the secondaries, including electrons and positrons might also escape, creating a funnel in the Galactic disk which allows them to flow out (see \cite{Diehl06,Diehl11,Diehl13,Siegert16b}). The total positron production in a large region around the Galactic Center corresponds to a power on the order of $10^{37.1} \, {\rm erg/s}$. This is $10^{-4.6}$ of the maximal energetic particle flow, of order $10^{51} \, {\rm  erg}$ in about $10^{9.3} \, {\rm s}$ (see above) even for a single massive star SN event, suggesting that much of the energy is vented out to the Galactic halo. Even allowing for a reduction by about a factor of 100, to account for the difference in CR electron fluxes from CR proton and Helium, would still leave a factor of $10^{-2.6}$. The contribution from the Galactic Center BH seems to be less than that which any possible surrounding sources could contribute.

In the balance between production of anti-protons from p - p collisions, as well as ${\rm \bar{{\rm p}}}$ - ${\rm \bar{{\rm p}}}$ collisions, the annihilation process p - ${\rm \bar{{\rm p}}}$ dominates. Those interactions will limit not only the ${\rm \bar{{\rm p}}}$ net production, but will also produce large numbers of neutrinos. On the other hand, the ${\rm p}$ vs ${\rm {\bar{p}}}$ interaction decreases with energy, while the ${\rm p}$ vs ${\rm p}$ interaction cross-section to produce ${\rm p}$ - ${\rm {\bar{p}}}$ pairs, rises with energy. These neutrinos will be crudely commensurate with the Poynting flux energy flow. They, however, could exceed the Poynting flux, if the production and immediate destruction of ${\rm \bar{{\rm p}}}$ greatly exceed the rate of accretion of ${\rm \bar{{\rm p}}}$, as this runs with the ratio of the cross sections.  Consequently, this process could emit a significant fraction of the rotational energy of the BH via neutrinos. 

We consider the following reactions: first for creating and annihilating anti-protons; here we include the primary protons. Note that these densities represent integrals over the momentum distribution, and the cross-sections include weighting due to the momentum phase-space distribution:

1) $${\rm p \, + \, p \, \to \,  p \, + \, p \, + \, p \, + \bar{p}}$$ with cross section $$\sigma_{pr,{\rm p} {\rm \bar{p}}} \, ;$$ protons have density $n_{\rm p}$ and anti-protons 
density $n_{\rm \bar{p}}$;

2) $${\rm p \, + \, \bar{p} \, \to \, {\rm multiple} \, \pi}$$ with cross-section $$\sigma_{de, {\rm p} {\rm \bar{p}}} \, .$$ -- The pions decay into neutrinos and other leptons.

3) The reaction $${\rm \bar{p} \, + \, \bar{p} \, \to \, \bar{p} \, + \, \bar{p} \, + \, p \, + \bar{p}}$$ has the same cross section as above for protons, $$\sigma_{pr,{\rm p} {\rm \bar{p}}}\, .$$

4) The production of anti-neutrons $${\rm \bar{p} \, + \, \bar{p} \, \to \, \bar{p} \, + \, \bar{n} \, + \bar{\pi}}$$ has the cross section  $$\sigma_{{\rm \bar {n}}, {\rm \bar{p}} {\rm \bar{p}}} \, .$$

There are corresponding analogous processes for producing or destroying protons.

The detailed balance equations are (adopting $c$ as an approximate typical velocity for the particles):

\begin{eqnarray}
\frac{d \, n_{\rm \bar{p}}}{d \, t} \, &=& \, \sigma_{pr,{\rm p} {\rm \bar{p}}} \, c \, n_{\rm p}^2 \, - \sigma_{de, {\rm p} {\rm \bar{p}}} \, c \, n_{\rm \bar{p}} \, n_{\rm p} \, + \, \sigma_{pr,{\rm p} {\rm \bar{p}}} \, c \, n_{\rm \bar{p}}^2 \, \nonumber \cr &-& \, \sigma_{{\rm \bar {n}}, {\rm \bar{p}} {\rm \bar{p}}} \, c \, n_{\rm \bar{p}}^2 \, - \, \frac{n_{\rm \bar{p}}}{\tau_{BH}} \, ,
\end{eqnarray}

\noindent and 

\begin{eqnarray}
\frac{d \, n_{\rm p}}{d \, t} \, &=& \, \sigma_{pr,{\rm p} {\rm \bar{p}}} \, c \, n_{\rm p}^2 \, - \sigma_{de, {\rm p} {\rm \bar{p}}} \, c \, n_{\rm \bar{p}} \, n_{\rm p} \, + \, \sigma_{pr,{\rm p} {\rm \bar{p}}} \, c \, n_{\rm \bar{p}}^2 \, \nonumber \cr &-& \, \sigma_{{\rm n}, {\rm p} {\rm p}} \, c \, n_{\rm p}^2 \, - \, \frac{n_{\rm p}}{\tau_{BH}} \, + \, \frac{n_{\rm p}}{\tau_{gal}} \, .
\end{eqnarray}

Here the last term in the previous equation, and the last two terms in this equation, represent accretion to the BH, and accretion from the outside, from an accretion disk for instance. Accretion from outside constitutes positive baryon number accretion. If many secondaries are created and accreted, their net baryon number is zero. Baryon number accretion derives from both populations.

Initially, we assume that the accretion terms are negligible. By virtue of particles and anti-particles behaving the same in corresponding cross-sections, we can now consider two situations:

First we consider the case, where $ n_{\rm \bar{p}} \, << \, n_{\rm p}$. In this case, the production of anti-protons via pair creation dominates, and for protons the reaction leading to neutron production dominates. So, in this case, the anti-protons grow in number, and the protons decrease in number. The situation is not stationary. 


Next, the condition of exact stationarity can be required, and expressions eq.(3) and eq.(4) can be subtracted from one another: By virtue of the symmetry of cross-sections between particles and anti-particles, the first three terms in eq.(10) are equal to the first thee terms in eq.(11), respectively, leaving the fourth term. This gives 

\begin{equation}
n_{\rm \bar{p}}^2 \,  \sigma_{{\rm \bar{n}}, {\rm \bar{p}} {\rm \bar{p}}} \, - \,  \sigma_{{\rm n}, {\rm p} {\rm p}} \, n_{\rm p}^2 \, = \, 0 \, .
\end{equation}

By virtue of the equivalence between particles and anti-particles, the two  cross-sections are identical and can be cancelled out. The result of the above operation is

\begin{equation}
n_{\rm \bar{p}}^2 \,  - \, n_{\rm p}^2 \, = \, 0 \, ,
\end{equation}

\noindent thus the density of protons and anti-protons is the same in stationarity, neglecting accretion both from outside and to the BH. This does not violate baryon number conservation since in this model, both protons and anti-protons are secondary; the baryon number is exactly zero.

It follows that the ratio of neutrino production via pion decay to ${\rm p}$ ${\rm \bar{p}}$ pair-production runs with the ratio of the two cross-sections, which is large; however, the cross-sections have to be weighted with the momentum phase space distribution as noted above. It follows that the time scale for refilling the momentum phase space necessary to yield large interaction rates is key to the effective neutrino luminosity. Correspondingly, the ratio of neutron production to ${\rm p}$ ${\rm \bar{p}}$ pair-production runs with the ratio of the two cross-sections, which is also large. The cross-section to make pions and ensuing neutrinos starts at small energy and is large, and so dominates over the neutron production. In this simplified picture creation and destruction balance, and so the momentum distribution adjusts itself to make the effective cross-sections match, moderated by the time scales of redistributing particles in momentum phase space.

Second, we allow for the accretion terms to be relevant. Then the difference of the two terms leads to

\begin{equation}
(n_{\rm p} \, - \, n_{\rm \bar{p}}) \, \left(\sigma_{{\rm n}, {\rm p} {\rm p}} \, (n_{\rm p} \, + \, n_{\rm \bar{p}}) + \frac{1}{\tau_{BH}} \right) \, = \, 
\frac{n_{\rm p}}{\tau_{disk}} \, .
\end{equation}

This means if the sum of the neutron production and the BH net accretion is much larger than the outside accretion (from, e.g., an accretion disk), then the relative difference

\begin{equation}
(n_{\rm p} \, - \, n_{\rm \bar{p}})/n_{\rm p}
\end{equation}

\noindent is small. The anti-proton density approaches the proton density. Next consider the sum of the two equations: A solution is possible, in which the creation of secondaries is mostly balanced by destruction, with some accreting to the BH, and an even smaller number providing net loss of particles to the outside.

The pion decay leading to neutrino production can be approximated well by the approach of \cite{Penrose71}, leading to an accretion of neutrinos to the BH. It also leads to a corresponding luminosity of outgoing neutrinos.

In summary, the test is clearly to determine the anti-proton fraction at the EeV energy scale. If that fraction is half of the sum of protons and anti-protons, then the neutrino luminosity is predicted to be large, with most neutrinos near GeV energies. We observe TeV energies in neutrinos, and above.

\subsection{The Penrose zones with magnetic fields}

All these arguments depend on the Penrose process \cite{Penrose71,Bardeen72}. However, the main difference to the collisional Penrose process (e.g. \cite{Bejger12,Hod16,Leiderschneider16,Schnittman18}) is that in our approach, based on the magnetic field observations, particles are scattered by magnetic field irregularities frequently and throughout the ergo-region. We can write the spectrum of magnetic field irregularities $I(k) k$ as energy density with wavenumber $k$, so that the mean free path can be written as 

\begin{equation}
r_g \, \frac{B^2/\{8 \, \pi\}}{I(k) \, k}
\end{equation}

where $r_g$ is the Larmor radius of the motion of a charged particle. This mean free path is far smaller then the scale of the ergo-region except for the very highest particle energies, spanning more than 9 orders of magnitude (from the values of $B \, \times \, r)$ observed, as discussed above and in \cite{ASR18,Gal19}.

Here we focus on the angular momentum transport and work out, how frequently the data show that the Penrose process happens; however, first we have to comment on orbits of particles versus the local 3D momentum phase space distribution:

\subsubsection{Momentum phase space distribution}

The near-BH region, the ergo-region (also referred to as the ergo-sphere, but is never actually anything near spherical) and its immediate outer environment, is full of a strong magnetic field (near $10^{10}$ Gauss for a ten Solar mass BH, as observed \cite{ASR18}), with a full spectrum of irregularities $I(k)$: Therefore the charged particle momentum phase space distribution is highly an-isotropic, and includes locally an extension to all possible orbits to EeV energies, the maximum allowed by the magnetic field. Magnetic field scattering remixes the orbits continuously in the locally non-rotating frame \cite{Bardeen72}; the magnetic field and the particles at all energies refer to the rotation, and so carry angular momentum. Similar to stellar orbits in globular clusters \cite{King66}, where the orbits are cut off by tidal forces, the phase space distribution cuts off where plunge orbits take all particles away. This is also akin to the loss-cone process \cite{Hills75} where stars are taken out of the distribution by going straight into a BH. So the angular momentum transport is governed on the outside of the ergo-region by a region with the thickness of scrambling the orbits by magnetic fields, which governs the ejection of particles carrying angular momentum, and anchoring the magnetic fields; so the thickness is strongly dependent on particle energy: we call this the outer Penrose zone: This consideration gives the angular momentum loss of the BH together with the ergo-region. The angular momentum transport on the inside of the ergo-region is governed by ubiquitous particle interaction, producing secondary protons and anti-protons with many more pions of either charge. The orbits are also scrambled in this zone by magnetic fields, but also by the new production of secondaries. Many of those particles going into the black hole carry less specific angular momentum than the BH itself \cite{Bardeen72}, and so take angular momentum net from the BH. We dub this the inner Penrose zone. The balance between loss towards the outside in the outer Penrose zone and loss to the inside, the BH, on the inside in the inner Penrose zone gives the net angular momentum loss of the BH. At the highest particle energies the outer and inner Penrose zones might touch.  Since the transport in this concept is given by secondary particles, the net transport to the outside is visible in the magnetic fields \cite{ASR18} and also in electron-positron pairs \cite{Siegert16a}, and, we posit, in pop 3$^{*}$ of \cite{Gaisser13}, which in this concept should carry an about equal number of anti-protons and protons. We note that all jets carry an electric current, driven by a proton-anti-proton pair plasma with a spectrum of $E^{-2}$ \cite{Gopal24} to EeV energies, which we identify here with this CR population, steepened by an ISM Kolmogorov spectrum of magnetic irregularities in the Galactic disk, so 1/3. Variable jets drive an electric field, which upon discharge drives particle energies much higher \cite{Gopal24}. Such discharges have been seen ubiquitously as synchrotron radio filaments \cite{Gopal24,Zadeh22}. One prediction in our model is that pop 3$^{*}$ of \cite{Gaisser13} should be composed by an equal number of protons and anti-protons, and this may be detectable  around and above TeV energies.

In \cite{Bardeen72} their Fig. 3 shows what fraction of velocity phase space - there simplified to equatorial orbits, so planar orbits - goes down into the BH. Because of the scrambling of charged particle orbits by the relatively strong magnetic fields there are in reality no orbits from or to infinity, within an interaction length of the horizon only orbits that either remain in the ergo-region or plunge down into the BH. Further inside the ergo-region all orbits are such that the particles remain in the ergo-region.  So as soon as magnetic scattering or new particle creation by collisions puts an orbit into the plunge region of momentum phase space that particle is directly lost. Since this part of phase space is not generally a cone, instead of a "loss-cone" we refer to it as the "plunge region of momentum phase space". That plunge region of momentum phase space exists only within an interaction length of the horizon, see the equation above, accounting both for magnetic scattering or particle collisions with creation of new particles.  The magnetic scattering interaction length is rigidity dependent, depending on the Larmor radius scaling linearly with rigidity, and the spectrum of resonant irregularities $I(k)$. For a spectrum of $I(k) \, \sim\, k^{-\beta}$, this gives an interaction length scaling with the power of $2 \, - \, \beta$. For a Kolmogorov spectrum this gives an interaction scaling with rigidity to a 1/3 power, lightning dominated turbulence gives a 5/3 power, while shock dominated turbulence gives an interaction length independent of rigidity \cite{Allen24}.  Collisions of particles to create new particles, such as lots of pions, or proton-anti-proton pairs, produces the most particles on such an orbit. As both $B^{r}$ and $B^{\phi}$, the observed components, scale as $\Delta^{1/2}$, the Larmor radius diverges near the horizon, and so the scattering by the magnetic field is weakened; on the other hand, the charged particle density also diverges near the horizon, both clear from the expressions above. For the collision rate between protons with other protons, including secondary protons, to be faster than pion decay implies extraordinary densities, of order $10^{24} \, {\rm cm^{-3}}$ or higher, easily possible with the expression above for the charged particle density.  This means that near the horizon the injection of mostly new particles into the plunge orbit part of momentum phase space dominates over pure magnetic field scattering. This has been the main thrust here, that secondary particles go onto plunge orbits, and so determine the spin-down.

\begin{figure}[htpb]
\centering
\includegraphics[bb=0cm 0cm 21.0cm 29.7cm,viewport=2.5cm 10.5cm 19.cm 19.0cm,clip,scale=1.0]{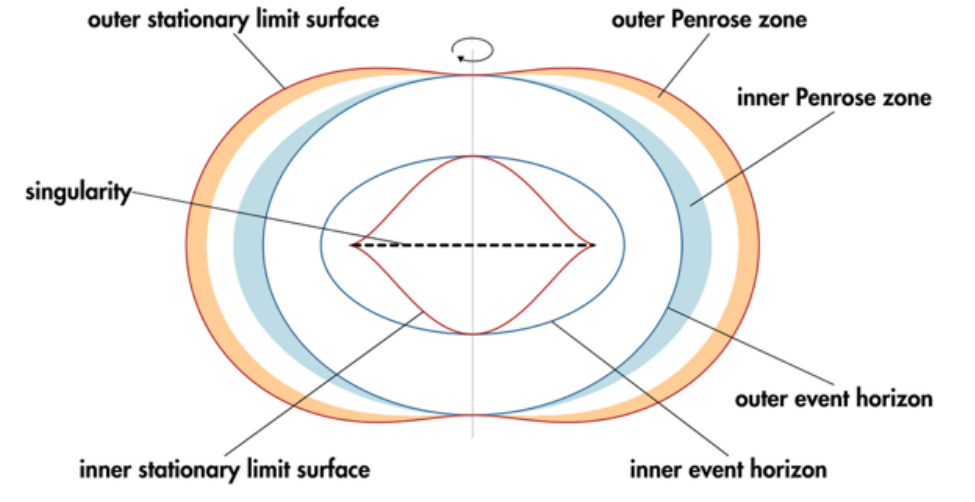}
\caption{The key elements of the Kerr black hole with rotation parameter $a=0.95$ are represented in a planar section containing the axis of rotation. The infinite curvature ring singularity  appears from lateral side view as a segment. This is hidden inside a structure of two horizons, the outer horizon being the boundary of the Kerr black hole. Two stationary limit surfaces (where $g_{tt}=0$) are positioned inside the inner horizon and outside the outer horizon, respectively. At the outer stationary limit surface the redshift is infinite and  photons cannot counterrotate, while inside it they will always corotate, similarly to all the other particles, irrespective of their initial direction. The ergo-region, lying between the outer stationary limit surface and the outer event horizon, contains the outer and inner Penrose zones, attached to these limiting surfaces.}
\end{figure}
\label{PenroseZones}

\subsubsection{Frequency of the Penrose process}

In pure spin-down angular momentum transport provides the main constraints:

The data show that the quantity $(B \, \times \, r)$ has the value $10^{16 \pm 0.12} \, {\rm Gauss \, \times \, cm}$ for both red supergiant and blue super-giant RSNe \cite{ASR18,Gal19}; this value is consistent with the numbers for super-massive black holes \cite{EHT19L5}. Using eq. 9 of \cite{Weber67} this corresponds, as seen from afar, to an angular momentum transport of $10^{38.4 \pm 0.24} \, {\rm \{erg \, s\}/s}$ for a 10 Solar mass BH, and using the assumption, that at the outer radius of the ergo-region (at $10^{6.4} \, {\rm cm}$ on the equator) the radial magnetic field is equal to the tangential field. This is the angular momentum transport just via the magnetic field. This angular momentum transport is enhanced by thermal and non-thermal particles, and similar to the ISM we assume here that non-thermal particles give the same angular momentum transport as the pure magnetic field, and the thermal particles give the same as this sum, the magnetic field and non-thermal particles added together: This gives a factor of $f_{ISM} \, = \, 4$ over the pure magnetic field case, for a final angular momentum transport of $10^{39.0 \pm 0.24} \, {\rm \{erg \, s\}/s}$. This number has to consistent with what particles transport.

How does this compare with what is carried by particles, also seen from afar?  The argument starts with protons and anti-protons, so $10^{-23.8} \, {\rm g}$, at $10^{6.4} \, {\rm cm}$ with close to the velocity of light $c$, so yielding $10^{- 6.9} \, {\rm g \, cm^2 \, s^{-1}}$ as a base unit of angular momentum. Extending the spectrum to EeV energies gives for an $E^{-2}$ spectrum \cite{Gopal24} a factor of the natural log of the range, so about $f_{CR} \, = \, 10^{1.2}$. Considering that pions result energetically 30 times as often from p-p collisions as proton-anti-proton pairs adds another factor of $f_{\pi's} \, = \, 10^{1.5}$ for a total angular momentum of $10^{- 4.2} \, {\rm g \, cm^2 \, s^{-1}}$. We will normalize these three factors to their nominal values, and the write $f_{ISM} \, = \, 4 \, f_{ISM, 0.6}$, $f_{CR} \, = \, 10^{1.2} \, f_{CR, 1.2}$, and $f_{\pi's} \, = \, 10^{1.5} \, f_{\pi's, 1.5}$.  So $10^{- 4.2} \, {\rm g \, cm^2 \, s^{-1}}$ gets a factor of $f_{CR, 1.2} \,f_{\pi's, 1.5}$.  What time scale per such step is required to match the observed angular momentum transport? The implied time scale is the Planck time of $\tau_{Pl} \, = \, 10^{-43.3} \, {\rm s}$ \cite{Planck1900}, which yields here $10^{39.1} \, {\rm \{erg \, s\}/s}$, consistent with the number indicated by magnetic field observations, as derived above. This says, that the Penrose process happens most efficiently for an $E^{-2}$ spectrum (also required for the electric current, and the large Debye length \cite{Gopal24}), and equally for each log bin of particle energy in the particle spectrum. It also says, that the Penrose process happens for a BH of any mass at near maximal rotation about $10^{46}$ times per second in terms of protons/anti-protons, and an order magnitude more often in terms of pions. This relies solely on the production of secondaries via collisions, and no accretion from far outside.

On the basis of observations discussed above we derive therefore the relationship 

\begin{equation}
(B \, \times \, r)^2 \, = \, \frac{f_{CR} \, f_{\pi's}}{f_{ISM}} \, \frac{m_p \, c}{\tau_{Pl}}
\end{equation}

\noindent for a BH of any mass in nearly maximal rotation, and in pure spin-down, so without any accretion.

The outer radius of the ergo-region drops out, and so this relationship becomes independent of proximity to the BH, as long as the scale is outside the ergo-region. The term with the factors $f_{CR}$, $f_{\pi's}$, and $f_{ISM}$, perhaps by coincidence, approximately equals $\{\hbar \, c\}/e^2$. The observations leading to this relationship range from a few $M_{\odot}$, to about $10^{10} \, M_{\odot}$.

We emphasize that in this interpretation radio observations of the magnetic field close to what we have proposed are near maximally rotating black holes, require the Planck time to match with protons/anti-protons and pions in angular momentum transport. This interpretation allows to understand the strength of the magnetic field; the magnetic field is determined by this process. This argument is valid for any black hole in near maximal rotation, and without any accretion.

This leads to the question, whether this can be thought of as spontaneous emission of a black hole in the sense of Einstein \cite{Einstein1917} and Feynman (\cite{Feynman1963}: Feynman Lectures of Physics, vol. I, p. 42.9). And if so, what qualifies as stimulated emission (see \cite{Falcke95,Gopal24})? The magnetic field in terms of $(B \times r)$ is larger by the square-root of the ratio of the power of the source to the minimum power implied here (also observed \cite{Punsly11,ASR18,Gal19,EHT19L5}. Therefore also in that case the Planck time is used.

\subsubsection{Circular orbits in the inner Penrose zone}

The lowest energies correspond to locally circular orbits in the inner Penrose zone \cite{Bardeen72}. This zone is governed predominantly by the numerous pions and their decay products; secondary protons and anti-protons feed the acceleration to the maximal energy allowed, but are way down in number. As pion production is energetically about 30 times proton-anti-proton pair production, and pions have about 1/10 the rest mass of protons/anti-protons, it entails that pions are about 300 times as numerous as protons/anti-protons if produced sufficiently fast. Neutrinos escape, but electron/positrons are trapped by the magnetic fields. They lose energy rather quickly, but can also be accelerated again in the bath of many waves. We can derive this temperature crudely as follows: Charged particles are easily thermalized in any post-shock region: if the equation of state is relativistic then the speed of sound is given by $c_{rel}^2 \, = \, c^2/3$, so that the typical velocities are some fraction of the speed of light, post-shock easily $c/3$, which for pions corresponds to order 30 MeV. Basically pions dominate the thermodynamics despite their short life-time. This requires that all time scales, like for producing pions, must be faster, and the densities correspondingly high. the conditions that p-p collisions to make proton-anti-proton pairs are faster than pion decay requires densities above $10^{24.5} $ per cc, a charged particle density plausible close to the horizon by the expression above.

\subsubsection{Observational tests}

This argument, that requires the Planck time, is derived from radio observations and their interpretation.

A priori we do not know, how many of the secondary particles are released to the outside, but in the interpretation, that the pop 3$^{*}$ of \cite{Gaisser13} and the particles driving an electric current in jets \cite{Gopal24} corresponds to the ejection of secondary protons and anti-protons from the Penrose zones around young stellar mass BHs, the strongest prediction to test is that AMS may be capable of determining these anti-protons and protons near to and beyond TeV. Annihilation of protons and anti-protons may also be detectable.

The electron-positrons detected by \cite{Martin10,Prantzos11,Siegert16a,Siegert16b,Prantzos17,MeraEvans22} may correspond to just the population derived from pion production and decay.

\section{Conclusions}

In the scenarios proposed here, we predict anti-protons to be seen above TeV energies \cite{AMS16} with the EeV proton component detected in fits of the cosmic ray data in \cite{Gaisser13}, \cite{Thoudam16}, and \cite{Auger20a}. These concepts lead us to a number of predictions and inferences:

\begin{itemize}

\item{} Massive stars, commonly found in multi-star systems, lose {\it orbital} angular momentum through magnetic winds.

\item{} This, in turn, allows a tightening of the binary system, and by tidal locking to an increase of rotation. Alternatively the core of the nascent star may rotate fast and remain in fast rotation during its rapid evolution.

\item{} Resulting BHs rotate initially near the maximum allowed value. This phase of high rotation is short-lived.

\item{} RSNe of former Red Super Giant stars and Blue Super Giant stars can be interpreted as winds emanating from the direct environment of the ergo-region of a BH, which rotates near the maximum allowed value.

\item{} The constancy of the value of the quantity $(B \, \times \, r)$, being independent of BH mass, in RSNe shows that protons can attain EeV energies.

\item{} The quantity $(B \, \times \, r)$  gives an angular momentum loss time scale of the BH of $\sim \, 10^{3.7} \, {\rm yrs} \, (M_{BH}/M_{\odot})$, so is proportional to the mass of the BH, here scaled to one Solar mass. For super-massive BHs we obtain the same value of the quantity $(B \, \times \, r)$, directly from M87 observations \cite{EHT19L5}, and indirectly from the minimum power observed \cite{Punsly11}. The time scale of angular momentum loss exceeds the age of the universe for any such BH of mass larger than $10^{6.5} \, M_{\odot}$, assuming it started at near maximal rotation. This value is remarkably close to the mass of our Galactic Center BH \cite{EHT19L5}. It follows that without spin-up intermediate mass BH are expected to rotate slowly \cite{Fuller22}.

\item{} This quantity leads to a power outflow of $\sim \, 10^{42.8} \, {\rm erg/s}$, independent of BH mass. This is seen for low power radio galaxies \cite{Punsly11}. For stellar mass BHs this is far above the Eddington power.

\item{} This power outflow comes purely from spin-down \cite{BZ77}, and is thus a minimum, matching observations of radio-quasars \cite{Punsly11}. 

\item{} The wind emanating from the ergo-region injects a CR population with an observed spectrum of $E^{-7/3}$ (due to transport out of the Galaxy, pop 3$^{*}$ in \cite{Gaisser13}; Table 3) and a maximum energy at EeV level. This population is predicted to show a fraction of anti-protons, half. At such a high charged particle density as required to make anti-protons, all higher mass nuclei will be destroyed by spallation; this component is only protons and anti-protons in our proposal.  This directly matches the argument about electric currents in jets being driven by a proton-anti-proton plasma with a spectrum of $E^{-2}$ \cite{Gopal24}. This is in addition to the stronger CR flux of all elements which is produced by SN-shocks (pop 1, 2 and 3 in \cite{Gaisser13}, Table 2). This destruction of heavier nuclei is actually a consistency check of our model, since the Gaisser et al. model \cite{Gaisser13} does not show such a heavy nuclei component, with this spectrum $E^{-7/3}$.

\item{} This model provides a floor to the anti-proton spectrum seen by AMS and limits determined by HAWC \cite{AMS16,HAWC18}  in the range of GeV to TeV. A consequence is that this component of the anti-protons should show a straight spectrum from near TeV energies all the way to EeV energies, with a $E^{-7/3}$ power law throughout.

\item{} The model suggests that in the ergo-region there is a cascading, collisional production of energetic particles, producing an abundance of secondaries. An electron/positron plasma is a primary product from these collisions. These secondaries produce strong drift currents, and exchange energy and angular momentum with the magnetic field \cite{Gopal24}.

\item{} The cascading might lead to a much higher production of anti-protons and protons than the number of protons actually accreted from far outside. Most of the anti-protons get annihilated in collisions with protons. In such a reaction, large numbers of neutrinos are produced, and those which escape can remove angular momentum. This could lead to an efficient reduction of rotational energy of the BH. This is possibly detectable as neutrinos with energies near GeV.

\item{} This scenario can be connected to a concept of inner and outer Penrose zones in the ergo-region. The observed numbers for the magnetic field imply the Planck time as the governing time scale: A BH rotating near maximum can accept a proton of low specific angular momentum per log bin of energy with the associated pions every Planck time.

\end{itemize}

\section{Acknowledgements}

PLB wishes to thank Nasser Barghouty, Susanne Blex, Julia Becker Tjus, Silke Britzen, Roland Diehl, Matthias Kaminski, Wolfgang Kundt, Norma Sanchez, and Gary Webb for stimulating discussions, as well as Carola Dobrigkeit, Roger Clay, and Roland Diehl and several others for helpful comments on the manuscript. The two coauthors Ben Harms and Phil Kronberg passed away during the work for this manuscript, but were integral in the discussions for decades leading up to it. The comments by the referees were extremely helpful in sharpening the argument, and adding further references.




\end{document}